\documentclass[fleqn,usenatbib]{mnras}

\usepackage{newtxtext,newtxmath}

\usepackage[T1]{fontenc}

\DeclareRobustCommand{\VAN}[3]{#2}
\let\VANthebibliography\thebibliography
\def\thebibliography{\DeclareRobustCommand{\VAN}[3]{##3}\VANthebibliography}

\usepackage{graphicx}	
\usepackage{amsmath}	
\usepackage{textcomp}	
\usepackage[dvipsnames]{xcolor}

\newcommand{\cs}{c_\mathrm{s}}
\newcommand{\Mp}{M_\mathrm{p}}
\newcommand{\Mth}{M_\mathrm{th}}
\newcommand{\hp}{h_\mathrm{p}}
\newcommand{\Hp}{H_\mathrm{p}}
\newcommand{\Rp}{R_\mathrm{p}}
\newcommand{\de}{\mathrm{d}}
\newcommand{\M}{\mathcal{M}}
\newcommand{\pd}[2]{\frac{\partial #1}{\partial #2}}
\newcommand{\td}[2]{\frac{\de #1}{\de #2}}

\newcommand{\const}{\mathrm{const.}}
\newcommand{\vect}[1]{\mathbf{#1}}

\newcommand{\changed}[1]{#1}

\newcommand\ba{\begin{eqnarray}}
\newcommand\ea{\end{eqnarray}}

\defcitealias{Goodman2001}{GR01}
\defcitealias{Rafikov2002}{RR02}

\newcommand\cmtrr[1]{{\color{red}[RR: #1]}}

\title{Planet-driven density waves in protoplanetary discs: numerical verification of nonlinear evolution theory}

\author[N. P. Cimerman and R. R. Rafikov]{
Nicolas P. Cimerman,$^{1}$\thanks{E-mail: npcphys@gmail.com (NPC)}
Roman R. Rafikov$^{1,2}\thanks{John N. Bahcall Fellow at the IAS}$
\\
$^{1}$Department of Applied Mathematics and Theoretical Physics, University of Cambridge, Wilberforce Road, Cambridge CB3 0WA, UK\\
$^{2}$Institute for Advanced Study, Einstein Drive, Princeton, NJ 08540, USA
}

\date{Accepted 2021 September 11. Received 2021 September 7; in original form 2021 August 7}

\pubyear{2021}

\begin{document}
\label{firstpage}
\pagerange{\pageref{firstpage}--\pageref{lastpage}}
\maketitle

\begin{abstract}
Gravitational coupling between protoplanetary discs and planets embedded in them leads to the emergence of spiral density waves, which evolve into shocks as they propagate through the disc. We explore the performance of a semi-analytical framework for describing the nonlinear evolution of the global planet-driven density waves, focusing on the low planet mass regime (below the so-called thermal mass). We show that this framework accurately captures the (quasi-)self-similar evolution of the wave properties expressed in terms of properly rescaled variables, provided that certain theoretical inputs are calibrated using numerical simulations (an approximate, first principles calculation of the wave evolution based on the inviscid Burgers equation is in qualitative agreement with simulations but overpredicts wave damping at the quantitative level). We provide fitting formulae for such inputs, in particular, the strength and global shape of the planet-driven shock accounting for nonlinear effects.
We use this nonlinear framework to theoretically compute vortensity production in the disc by the global spiral shock and numerically verify the accuracy of this calculation. Our results can be used for interpreting observations of spiral features in discs, kinematic signatures of embedded planets in CO line emission ("kinks"), and for understanding the emergence of planet-driven vortices in protoplanetary discs.
\end{abstract}





\section{Introduction}
\label{sect:intro}


Gravitational coupling of young planets with the protoplanetary discs in which they form is known to give rise to global spiral density waves. These planet-induced waves may be responsible for the non-axisymmetric structures observed around multiple disc-hosting stars, for example spiral patterns seen in scattered light observations of MWC 758 \citep{Grady2013, Benisty2015}, SAO 206462 \citep{Muto2012,Garufi2013}, and other systems.

Density waves launched by planets carry angular momentum and energy, which can be deposited into the disc fluid giving rise to disc evolution \citep{Goodman2001,Rafikov2002,Rafikov2016} and gap formation \citep{Lin1993,Rafikov2002II}, provided that the wave can be damped either linearly \citep{Takeuchi1996,Miranda2020I,Miranda2020II} or nonlinearly \citep{Goodman2001,Rafikov2002}. Nonlinear wave dissipation naturally results from nonlinear steepening of the  wave profile and its eventual evolution into a shock, even for low wave amplitudes. Irreversible energy dissipation at the shock is the ultimate cause of the nonlinear damping of a density wave \citep{Goodman2001,Rafikov2016}.

The details of propagation and evolution (damping) of weakly nonlinear planet-driven waves have been studied in the local (homogeneous shearing sheet) approximation by \citet{Goodman2001} (hereafter \citetalias{Goodman2001}). They showed that for low mass planets, $\Mp\lesssim \Mth$, the problem of linear wave excitation by planetary gravity can be naturally separated from the subsequent wave propagation affected by nonlinear effects, and explored both stages. Here $\Mth$ is the characteristic mass scale, so called {\it thermal mass}
\ba  
\Mth=\frac{\cs^3}{\Omega G}=\left(\frac{\Hp}{\Rp}\right)^3 M_\star,
\label{eq:Mth}
\ea  
(where $\cs$ is the sound speed, $\Omega$ is the orbital angular frequency, $\Hp$ is the disc scale height at the planetary distance $\Rp$, and $M_\star$ is the stellar mass) such that planets with $\Mp\sim\Mth$ launch density waves which are nonlinear already at excitation. Subsequently \citet{Dong2011I,Dong2011II} carried out high resolution hydrodynamical simulations of planet-launched density waves in the shearing sheet approximation. They verified the main results of the \citetalias{Goodman2001} theory, in particular the prediction for the distance from the planet $l_{\rm sh}$ at which the wave shocks (see also \citealt{Yu2010}), and the evolution of the amplitude and width of the wave profile in the asymptotic "$N$-wave" regime. Additionally, they investigated the production of vortensity $\Delta \zeta$ by the planet-driven spiral shocks, showing it to be a steep function of planet mass ($\Delta \zeta \propto \Mp^3$). 

The local theory of \citetalias{Goodman2001} has been subsequently extended to global discs in \citet{Rafikov2002} (hereafter \citetalias{Rafikov2002}), fully accounting for radial gradients of the protoplanetary disc properties (e.g. $\cs$ and gas surface density $\Sigma$) and curvature effects. This more general, global framework has in particular allowed \citet{Rafikov2002II} to explore gap opening by migrating planets accounting for the non-locality of nonlinear wave damping, which was later verified numerically in \citet{Li2009} and \citet{Yu2010}. 

The goal of our present work is to quantitatively verify the global theoretical framework of \citetalias{Rafikov2002} using hydrodynamical
simulations. We study various aspects of excitation, propagation and decay of the density waves driven by sub-thermal mass ($\Mp\lesssim \Mth$) planets. Also, following \citet{Dong2011II}, we use the global theoretical framework of \citetalias{Rafikov2002} to compute vortensity generation by global planet-driven spiral shocks and verify these predictions numerically. There are several key motivations for this study. 

First, there has been limited amount of past work trying to verify nonlinear propagation of global density waves, mainly documenting the evolution of the wave profile towards the asymptotic $N$-wave regime \citep{Duffell2012} and examining the deviation of the spiral shape from the linear prediction \citep[see \S\ref{sec:phishock}]{Zhu2015}. Here we aim to provide a systematic study of the wave excitation and evolution, covering a broad range of the relevant parameters --- disc aspect ratio $\hp=\Hp/\Rp$, planet mass $\Mp$, and disc surface density profile $\Sigma(R)$ --- using a variety of diagnostics.  

Second, theoretical frameworks of \citetalias{Goodman2001} and \citetalias{Rafikov2002} reduce the full set of fluid equations to a single inviscid Burgers equation in the limit of a weakly nonlinear density wave. So far the accuracy of this approximation (which has been recently employed in \citealt{Bollati2021}) has not been studied, and we provide its systematic test in this work. 

Third, the global theory of \citetalias{Rafikov2002} explicitly assumes that a linear effect --- excitation of a single-armed spiral wave \citep{OgilvieLubow2002} by the planetary potential --- takes place only close to the planet, and that far from it only nonlinear effects regulate wave evolution. However, recent studies \citep{Bae2018,Miranda2019I} have shown that linear effects (in the form of evolving interference of the individual modes comprising the wave profile) continue driving wave evolution in the disc interior to the planetary orbit even far from the planet, leading to
\changed{
an evolution of a single-armed density wave into multiple arms.
}
Our current work will examine how this effect modifies the picture of global density wave propagation in discs. 

This paper is structured as follows. In  \S\ref{sec:theory}, we describe our physical setup and summarize the key ingredients of the global wave evolution theory of \citetalias{Rafikov2002}. Upon describing our numerical tools in \S\ref{sect:numerics-gen}, we present results on the nonlinear evolution of the global planet-driven density waves for different planet masses in \S\ref{sec:res}, as well as the fitting formulae for the resultant shock strength and shape in \S\ref{sec:shock}. In \S\ref{sec:resvort} and Appendix \ref{sec:thvort} we develop a framework for computing the vortensity production by the planet-driven spiral shock and verify it numerically. We explore the effect of varying disc parameters on wave evolution in \S\ref{sec:resdisc}. Our results are further discussed in \S\ref{sec:disc} and summarized in \S\ref{sec:sum}.


\section{Problem Setup and Theoretical background}
\label{sec:theory}


In this section we first describe the physical setup for the problem under consideration (\S\ref{sect:setup}), and then provide relevant theoretical background (\S\ref{sect:wave-lin},\ref{sect:wave-nonlin}).


\subsection{Problem Setup}
\label{sect:setup}


We consider a planet of mass $M_\mathrm{p}$ orbiting a central star of mass $M_\star$ on a circular orbit with a semi-major axis $R_\mathrm{p}$ that lies within a two-dimensional gas disc. The two-dimensional approximation is appropriate for thin discs, such that the disc aspect-ratio $h = H/R = \cs/(\Omega R) \ll 1$. Planetary gravity perturbs the motion of disc fluid, giving rise to a density wave that we explore in this work. We adopt polar coordinates $(R,\phi)$ to describe this problem. 

The background disc state, unperturbed by the planet, has a power-law profile of the surface density
\begin{align}
	\Sigma_0(R) =  \Sigma_\mathrm{p} \left(\frac{R}{R_\mathrm{p}}\right)^{-p},
	\label{eq:model-PLs}
\end{align}
where $p$ is a constant and $\Sigma_\mathrm{p}=\Sigma_0(R_\mathrm{p})$. Throughout this work we assume the mass of the disc to be small, $M_\mathrm{d} \ll M_\star$, such that its self-gravity can be neglected. 

For all our models, we adopt a {\it globally isothermal} equation of state (EoS), $P = \cs^2 \Sigma$, where $P$ is the vertically-integrated pressure, $\Sigma$ is the surface density, and $\cs$ is the spatially constant speed of sound. We opted to use this barotropic EoS instead of an often used nonbarotropic \textit{locally} isothermal EoS, for which the sound speed follows a prescribed radial profile $\cs(R)$, for several reasons. 

First, it has recently been shown by \citet{Miranda2019II,Miranda2020I,Miranda2020II}, that the use of a locally isothermal EoS leads to a non-conservation of angular momentum flux (AMF) of the density waves freely propagating through the disc even in the absence of explicit dissipation. On the other hand, a barotropic EoS (in particular, the globally isothermal EoS) conserves wave AMF after excitation \citep[see also][]{Lin2011,Lin2015}, which greatly simplifies our analysis of the problem (see below). Second, the globally isothermal assumption eliminates baroclinic vorticity driving, see  \S\ref{sect:EoS}. Third, the use of this EoS reduces the number of parameters characterizing the problem at hand.

The unperturbed background disc is in radial centrifugal balance, accounting for the radial pressure gradient: radial velocity $u_{R,0}(R) = 0$, azimuthal velocity $u_{\phi,0}(R) = R\Omega_0(R)$, where 
\begin{align}
	\Omega_0^2(R) = \Omega_\mathrm{K}^2(R) + \frac{\cs^2}{R \Sigma_0(R)} \td{\Sigma_0(R)}{R},
	\label{eq:radHSE}
\end{align}
and $\Omega_\mathrm{K} = \sqrt{GM_\star/R^3}$ is the Keplerian orbital frequency. 

Our fiducial disc model has an aspect-ratio at the planet's radius, $h_\mathrm{p} = 0.05$ and a surface density slope $p = 3/2$. The latter results in an initial vortensity profile that is almost constant for slightly sub-Keplerian discs, see \S\ref{sec:resvort}.

When presenting our results we adopt units where $G = M_\star = \Omega_K(R_\mathrm{p}) = \Sigma_0 (R_\mathrm{p}) = 1$.


\subsection{Linear evolution of planet-driven density waves}
\label{sect:wave-lin}


\begin{figure*}
\centering
\includegraphics[width=0.99\textwidth]{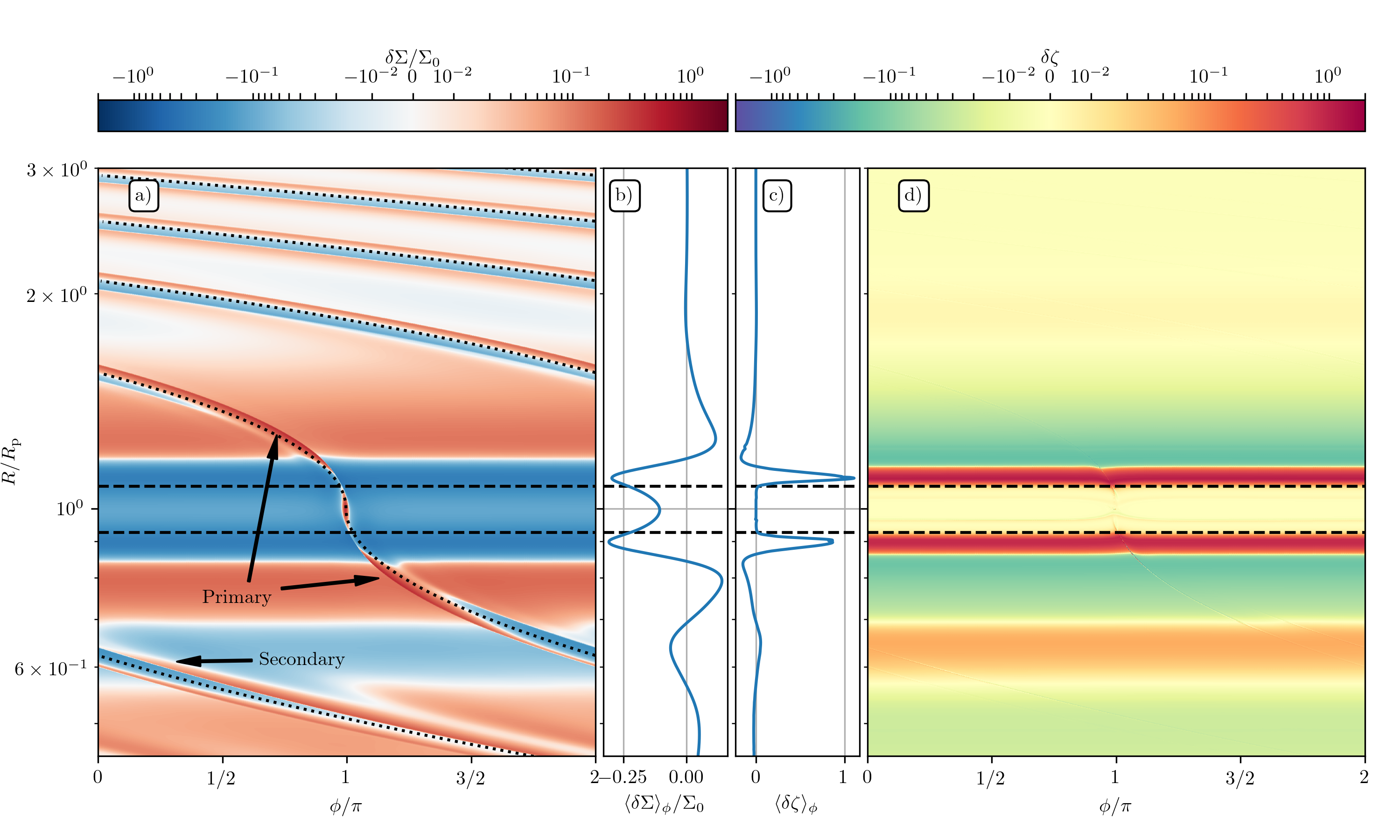}
\vspace*{-0.1cm}
\caption{A snapshot (at $t = 740$ planetary orbits) from one of our simulations using fiducial disc parameters and an intermediate-mass planet $M_\mathrm{p}/M_\mathrm{th} = 0.25$. (a) An $R-\phi$ map of the normalized surface density perturbation $\delta\Sigma/\Sigma_0(R)$. The locations of the primary and secondary spiral arms are indicated by arrows. The linear prediction (Eq. (\ref{eq:philin})) is shown by the black dotted line. (b) Radial profile of the azimuthally averaged surface density perturbation $\delta\Sigma/\Sigma_0(R)$. (c) Radial profile of the azimuthally averaged vortensity perturbation $\delta\zeta$. (d) Map of the vortensity perturbation $\delta\zeta$. The dashed horizontal lines show the predicted shock locations $R = R_\mathrm{p} \pm l_\mathrm{sh}$. This long-term simulation was run at the resolution $N_R \times N_\phi = 3448 \times 7200$.}
\label{fig:dvort_map}
\end{figure*}

The linear excitation of spiral density waves by massive perturbers has been extensively studied in the literature \citep[e.g.][]{Goldreich1980,Ward1997}.
Planetary gravity gives rise to the torque density per unit radius
\begin{align}
	\td{T(R)}{R} = - R \oint \Sigma(R,\phi) \pd{\Phi_\mathrm{p}(R,\phi)}{\phi} \,\de \phi,
	\label{eq:dTdR}
\end{align}
(here $\Phi_\mathrm{p}$ is the planetary potential) which imparts angular momentum flux on the density wave close to the planet, with the main contribution coming from a region $\lesssim 2\Hp$ away from the planet (\citetalias{Goodman2001}; \citealt{Dong2011I}). Farther away, the waves can be considered as freely propagating \citepalias{Goodman2001}.

In the absence of dissipation and in barotropic discs, freely propagating waves preserve their angular momentum flux (AMF) $F_J$ defined as 
\begin{align}
	F_J(R) &= R^2 \oint \Sigma(R,\phi) u_R(R,\phi) \delta u_\phi(R,\phi) \,\de \phi,
	\label{eq:FJ}
\end{align}
(with $\delta u_\phi(R,\phi)=u_\phi(R,\phi)-u_{\phi,0}(R)$), i.e. $\partial_R F_J=0$. For planet-driven waves, the characteristic amplitude of the AMF (one-sided Lindblad torque) is \citep{Goldreich1980}
\begin{align}
	F_{J,0} = \left(\frac{M_\mathrm{p}}{M_\star}\right)^2 h^{-3}_\mathrm{p} \Sigma_\mathrm{p} r^4_\mathrm{p} \Omega^2_\mathrm{p},
	\label{eq:FJ0}
\end{align}
and we will use it as a reference value. 

To zeroth order, the freely propagating planet-driven density wave appears as a narrow, single-armed spiral wrapped by differential rotation in the inner and outer disc, see Figure \ref{fig:dvort_map}a for illustration. A simple linear prediction for the azimuthal location of this spiral (\citetalias{Rafikov2002}; \citealt{OgilvieLubow2002}), based on phase coherence arguments for linear WKB wave modes, reads
\begin{align}
	\phi_\mathrm{lin}(R) = \phi_\mathrm{p} + \operatorname{sign} \left(R-R_\mathrm{p}\right) \int_{R_\mathrm{p}}^{R} \frac{\Omega\left(R^{\prime}\right)-\Omega_\mathrm{p}}{c_{0}\left(R^{\prime}\right)} \de R^{\prime}.
	\label{eq:phi_lin}
\end{align}
This relation gives an approximate shape of the curves traced out by the peak of the density perturbation in $R-\phi$ coordinate.  

However, it has been realized recently that a single-peak structure does not fully capture the shape of the spiral density wave in the inner disc \citep{Dong2015,Fung2015}. Instead, far enough from the planet, the wave \changed{ is comprised of
several narrowly spaced (for low $\Mp$) spiral arms, as indicated in Figure \ref{fig:dvort_map}a.
This redistribution of AMF was understood in \citet{Bae2018} and \citet{Miranda2019I} as a manifestation of {\it linear} evolution of the density waves in differentially rotating discs, following from their weakly dispersive nature.
}
It was shown that the interference of different azimuthal harmonics comprising the perturbation pattern evolves in the inner disc in such a way as to cause the appearance of secondary, tertiary, etc. arms after the wave has travelled far enough from the planet. During this process, the primary spiral arm steadily transfers some of its AMF to the secondary (and higher order) spiral(s), thereby decreasing in amplitude even in the absence of any dissipative effects (the full wave AMF is conserved in the absence of damping in barotropic discs). \citet{Arza2018} and \citet{Miranda2019I} have also shown that this phenomenon is not unique to waves driven by a planet, but occurs for any passively propagating density wave. In this work, we will explore how this linear effect impacts nonlinear wave evolution.


\subsection{Nonlinear evolution of global density waves}
\label{sect:wave-nonlin}


While the excitation of waves by planets can be described by linear theory, their propagation over large distances is subject to nonlinear effects, even for small wave amplitudes. This is the process that ultimately makes waves steepen, form a shock, and dissipate, thus depositing their angular momentum to the disc.

Keeping the most important nonlinear terms in their analysis, \citetalias{Goodman2001} have shown that the evolution of weakly nonlinear waves can be reduced to a one-dimensional nonlinear wave propagation problem described by the inviscid Burgers equation (see below). Their calculation was {\it local} as it adopted the shearing-sheet geometry and assumed a uniform background state of the disc. \citetalias{Rafikov2002} extended that analysis to the {\it global} case by allowing for an inhomogenous, radially-structured  disc and accounting for the cylindrical geometry, still reducing the problem of wave propagation to the same mathematical form. 

The basis of \citetalias{Rafikov2002} analysis lies in the special rescaling of variables: radius $R$ gets replaced with a time-like coordinate\footnote{We renamed the parameter $t$ from \citetalias{Rafikov2002} to $\tau$ to avoid confusion with time $t$, which we use e.g. in the hydrodynamic equations. We also added a sign function that changes from inside to outside the planet's orbit, such that Eq. (\ref{eq:Burger}) is valid in both regions. This is due to the characteristics changing roles at this point \citepalias[see][]{Goodman2001}.} $\tau$ defined as a function of $R$ as 
\begin{align}
\label{eq:tau_g}
\tau(R) &\equiv \operatorname{sign} (R-R_\mathrm{p}) \times \frac{3}{2} \frac{M_\mathrm{p}}{M_\mathrm{th}} \frac{R_\mathrm{p}}{H_\mathrm{p}} \int_{R_\mathrm{p}}^{R} \frac{\Omega\left(R^{\prime}\right)-\Omega_\mathrm{p}}{c_{0}\left(R^{\prime}\right) g\left(R^{\prime}\right)} \de R^{\prime},
\end{align}
where the auxiliary function $g(R)$ is defined by
\begin{align}
\label{eq:g_g}
g(R) &\equiv \frac{2^{1 / 4}}{R_\mathrm{p} c_\mathrm{p} \Sigma_\mathrm{p}^{1 / 2}}\left(\frac{R \Sigma_{0} c_{0}^{3}}{\left|\Omega-\Omega_\mathrm{p}\right|}\right)^{1 / 2},
\end{align}
while azimuthal angle $\phi$ gets replaced by a space-like coordinate $\eta$ defined as
\begin{align}
\label{eq:eta_g}
\eta(R,\phi) &\equiv \frac{3}{2} \frac{R_\mathrm{p}}{H_\mathrm{p}} \left[\phi - \operatorname{sign}\left(R-R_\mathrm{p}\right) \int_{R_\mathrm{p}}^{R} \frac{\Omega\left(R^{\prime}\right)-\Omega_\mathrm{p}}{c_{0}\left(R^{\prime}\right)} \de R^{\prime}\right].
\end{align}
Note that $\tau(R_\mathrm{p}) = 0$ and also $\eta(R,\phi)=0$ for $\phi=\phi_\mathrm{lin}(R)$, i.e. at the location of the wake in linear theory, see equation (\ref{eq:phi_lin}). Finally, the density perturbation $\Sigma-\Sigma_0$ gets rescaled to a new independent variable $\chi$ defined as
\begin{align}
\label{eq:chi_g}
\chi(R,\phi)= \chi(\tau,\eta)&\equiv \frac{\gamma+1}{2} \frac{M_\mathrm{th}}{M_\mathrm{p}} \frac{\Sigma-\Sigma_{0}}{\Sigma_{0}} g(R). 
\end{align}
In all these definitions $\Sigma_0$ and $c_0$ can in general be functions of $R$ (unlike \citetalias{Goodman2001}). 

With these new variables the free propagation of a weakly nonlinear density wave is described by the inviscid Burgers equation \citepalias{Rafikov2002}
\begin{align}
	\partial_\tau \chi - \chi \partial_\eta \chi = 0,
	\label{eq:Burger}
\end{align}
the same as in the local limit of \citetalias{Goodman2001}. If we neglect nonlinearity, i.e. drop the second, quadratic in $\chi$, term in this equation, then $\chi=\chi(\eta)$ becomes independent of $\tau$ and $R$. Then equation (\ref{eq:chi_g}) allows one to directly determine how the surface density perturbation $\Sigma-\Sigma_0\propto \Sigma_0(R)/g(R)$ varies due to AMF conservation in the course of linear wave propagation in a differentially rotating, non-uniform disc \citepalias{Rafikov2002}.

Making use of the WKB approximation, one can obtain the following expression\footnote{We discuss in \S\ref{sec:res} how well this approximation describes the full wave AMF $F_J$ given by the equation (\ref{eq:FJ}) in our simulations.} for the wave AMF in terms of $\chi$ \citepalias{Goodman2001,Rafikov2002}: 
\begin{align}
	F_J^\mathrm{WKB} (\tau) = \frac{\sqrt{2}}{3} \frac{c_\mathrm{p}^{3} R_{\mathrm{p}} \Sigma_{\mathrm{p}}}{\Omega_\mathrm{p}} \left(\frac{M_\mathrm{p}}{M_\mathrm{th}}\right)^2 \Phi\left(\tau \right).
	\label{eq:FJPhi}
\end{align}
Thus, the evolution of the AMF $F_J^\mathrm{WKB}$ is entirely dictated by the behavior of the integral
\begin{align}
	\Phi (\tau) = \int \chi^2(\eta,\tau) \, \de \eta.
	\label{eq:Phi_def}
\end{align}
In particular, in the absence of nonlinearity, when $\chi$ is independent of $\tau$, one finds that both $\Phi$ and $F_J^\mathrm{WKB}$ are conserved in the course of wave propagation, as expected.

We now summarize some key results of \citetalias{Goodman2001} and \citetalias{Rafikov2002} regarding propagation of weakly nonlinear planet-driven waves.

\begin{itemize}

\item In the Burgers equation framework (\ref{eq:tau_g})-(\ref{eq:Burger}) wave evolution is {\it self-similar}. In other words, equation (\ref{eq:Burger}) does not contain any physical parameters of the problem ($M_\mathrm{p}$, $\Sigma_0(R)$, $c_0(R)$, $\Omega_\mathrm{p}$, etc.), all of which are absorbed into the definitions of variables $\tau$, $\eta$, $\chi$ and initial conditions. 

\item For a weakly nonlinear wave excited by a planet with $M_\mathrm{p}\lesssim M_\mathrm{th}$ the regions of linear wave excitation (within $(1-2)H_\mathrm{p}$ from the planet) and its subsequent nonlinear propagation are spatially separated. 

\item As a result of nonlinear evolution, the wave inevitably shocks at some radial separation $l_\mathrm{sh}$ away from the planet. This shocking length is set by the value of $\tau=\tau_\mathrm{sh}$ at which the characteristics of the Burgers equation (\ref{eq:Burger}) first cross. Using the linear wake profile from the excitation region (determined by solving the linearized perturbation equations) as an initial condition for the Burgers equation, \citetalias{Goodman2001} found
\begin{align}
\label{eq:taush}
	\tau_\mathrm{sh} &\simeq \tau_0 + 0.53, ~~~~~{\rm with}\\
	\tau_0 &= 1.89 M_\mathrm{p}/M_\mathrm{th}
\label{eq:tau0}
\end{align}
being the point at which the initial conditions are applied. The radial distance from the planet, at which the wave starts to shock (i.e. where $\tau=\tau_\mathrm{sh}$) is
\begin{align}
\label{eq:lsh}
	l_\mathrm{sh} \simeq 0.8 H_\mathrm{p} \left( \frac{\gamma + 1}{12/5} \frac{M_\mathrm{p}}{M_\mathrm{th}} \right)^{-2/5}.
\end{align}
One can see that for $M_\mathrm{p}\lesssim M_\mathrm{th}$ the shocking length $l_\mathrm{sh}$ indeed lies outside the linear wave excitation region, $l_\mathrm{sh} \gtrsim H_\mathrm{p}$.

\item After shocking, the wave profile asymptotically (for $\tau\gtrsim \tau_\mathrm{sh}$) evolves into an $N$-wave shape \citep{LL,whitham2011linear}, which is a typical outcome of the wave evolution governed by the Burgers equation. In this regime, the azimuthal width of the wake grows as $\Delta \eta \propto \tau^{1/2}$ while the wave AMF decays as $\Phi \propto \tau^{-1/2}$, which results in $F_J^\mathrm{WKB} \propto \tau^{-1/2}$ \citepalias{Goodman2001}.

\end{itemize}

\begin{figure}
\centering
\includegraphics[width=0.49\textwidth]{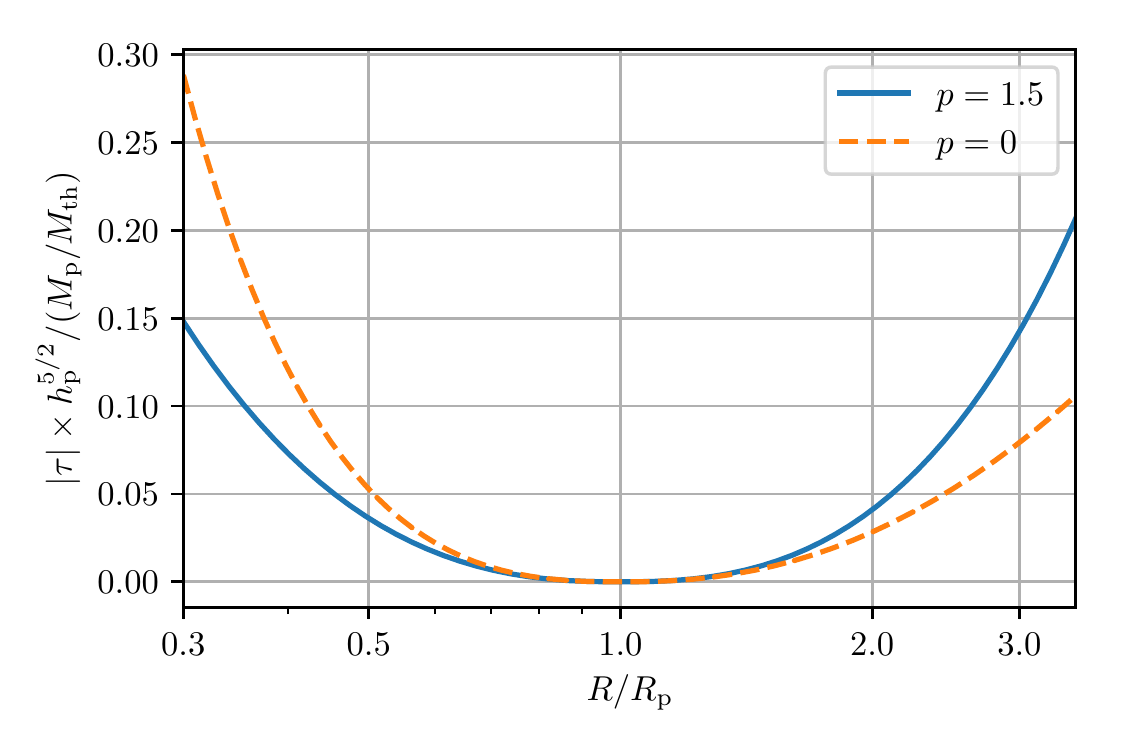}
\vspace*{-0.7cm}
\caption{Radial profile of the coordinate $\tau(R)$ for the disc profiles considered in this work, as labeled in the legend. The dependence on $h_\mathrm{p}^{5/2}$ and
$M_\mathrm{p}/M_\mathrm{th}$ is absorbed by rescaling the vertical axis. One can see that for a uniform disc $p=0$, nonlinearity is accelerated (slowed down) in the inner (outer) disc, as compared to the fiducal value ($p = 1.5$).}
\label{fig:tau-r}
\end{figure}

Assuming a Keplerian rotation profile and taking the globally isothermal limit ($c_0 = \cs = \const$, $\gamma \rightarrow 1$), the variables $\tau,\eta,\chi$ take the following form:\footnote{We use the assumption $\Omega(R) = \Omega_\mathrm{K}(R)$ in order to keep the analytical form of the coordinates simpler. We checked that accounting for sub-Keplerian rotation results in negligible
changes for thin discs.}
\begin{align}
	\nonumber
	\tau = \operatorname{sign} (R-R_\mathrm{p}) &\times \frac{3}{2^{5/4}} \frac{M_\mathrm{p}}{M_\mathrm{th}} h_\mathrm{p}^{-5/2} \\
	\color{black}	
	&\times \left\vert \int_{1}^{R/R_\mathrm{p}} \left\vert s^{3/2} - 1 \right\vert^{3/2} s^{p/2-11/4}\, \de s \right\vert,
	\label{eq:tau}
\end{align}
\begin{align}
	\eta &= \frac{3}{2 h_\mathrm{p}} \left[ \phi - \operatorname{sign} (R - R_\mathrm{p}) \frac{1}{h_\mathrm{p}}
	\left( 3 - 2\sqrt{\frac{R_\mathrm{p}}{R}} - \frac{R}{R_\mathrm{p}} \right) \right], 
	\label{eq:eta}\\
	\chi &= \frac{M_\mathrm{th}}{M_\mathrm{p}} \frac{\Sigma-\Sigma_{0}}{\Sigma_{0}}
	\left( \sqrt{2} h_\mathrm{p} \frac{\left(R/R_\mathrm{p}\right)^{-p+1}}{\left\vert \left(R/R_\mathrm{p}\right)^{-3/2} - 1 \right\vert}
 \right)^{1/2} 
 \label{eq:chishock},\\
 \phi_\mathrm{lin} &= \phi_\mathrm{p} + \operatorname{sign} (R - R_\mathrm{p}) \frac{1}{h_\mathrm{p}}
	\left( 3 - 2\sqrt{\frac{R_\mathrm{p}}{R}} - \frac{R}{R_\mathrm{p}} \right).
	\label{eq:philin}
\end{align}

The behaviour of $\tau$ for the two values of surface density slopes $p$ that we consider in this work is shown in Fig. \ref{fig:tau-r}. In a uniform disc ($p=0$) higher (lower) values of $\tau$ are reached in the inner (outer) disc, as compared
to the fiducal value $p = 1.5$, meaning that nonlinearity is accelerated (slowed down) in a uniform disc. Also, as can be seen from Eq. (\ref{eq:tau}), $\tau \propto h_\mathrm{p}^{-5/2}$, so that increasing the disc scale height would lead to slower nonlinear wave evolution over the same radial distance.


\section{Numerical Setup}
\label{sect:numerics-gen}


We now describe several different numerical tools used in this work. 


\subsection{Hydrodynamical simulations}
\label{sect:numrics}


We perform global, nonlinear hydrodynamic simulations of planet-disc interaction using Athena++\footnote{Athena++ is publicly available on \href{https://github.com/PrincetonUniversity/athena-public-version/}{GitHub}.} \citep{Athenapp2020}. The code solves the hydrodynamic equations in conservative form for mass and momentum using a Godunov scheme.
\begin{align}
    \pd{\rho}{t} + \nabla \cdot (\rho \vect{u}) &= 0, \\
    \pd{(\rho \vect{u})}{t} + \nabla \cdot (\rho \vect{u} \otimes \vect{u} + P \vect{I}) &= - \rho \nabla \Phi,
\end{align}
where $P$ is the gas pressure, $\vect{I}$ the identity tensor, and $\Phi$ is the gravitational potential.

In our globally isothermal setup, the energy equation is not solved.
Our fiducial setup computes the fluxes at cell-interfaces using second order accurate (linear) spacial reconstruction and Roe's approximate Riemann solver. We have performed additional tests using the HLLE solver and found no noticeable differences in our results. The equations are integrated in time using a second order Runge-Kutta scheme.

The total potential is given by $\Phi = \Phi_\star + \Phi_\mathrm{p},$ where the potential due to the central star is Newtonian $\Phi_\star = -GM_\star/R$ and the planet's potential is $\Phi_\mathrm{p}$. We neglect the indirect potential contribution that arises due to the acceleration of our non-inertial reference frame that is centred on the star. This is mainly to allow a direct comparison to the linear results of \citet{Miranda2019I}, who used the same approach\footnote{As noted by \citet{Miranda2019I}, the indirect term is linear in $R$, thus becoming more important for $R \gg R_\mathrm{p}$. It might be more significant when considering even larger domains of the disc.}. We performed tests that include the indirect term and they have shown no noticeable differences in the main results.

\begin{figure*}
\centering
\includegraphics[width=\textwidth]{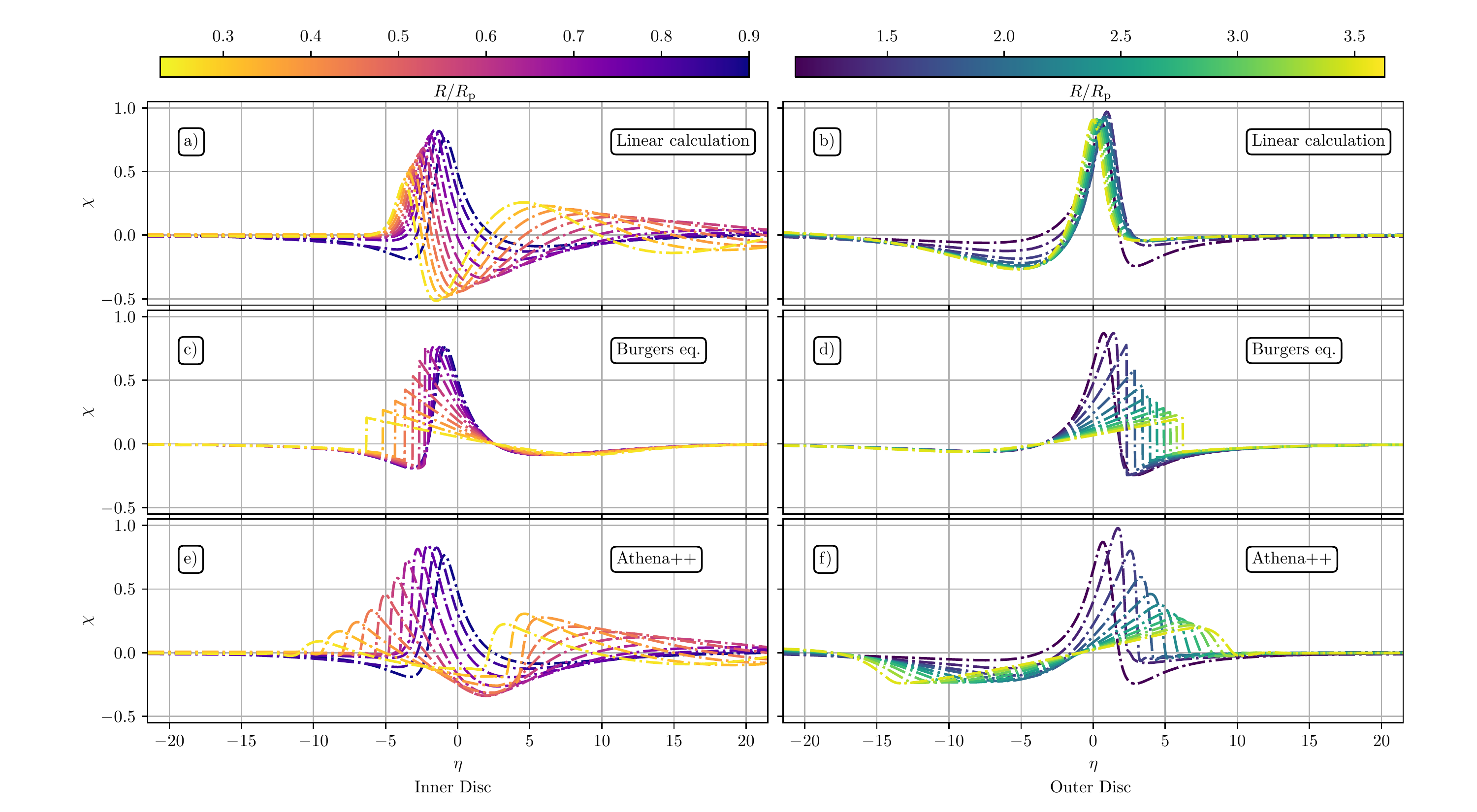}
\vspace*{-0.7cm}
\caption{Evolution of the wave profile for a low mass planet, $M_\mathrm{p} = 0.01 M_\mathrm{th}$. From top to bottom we show azimuthal slices (at several radii) of the planet-driven density perturbation obtained by (panels a-b) solving the linear perturbation equations, (panels c-d) solving Burgers equation and (panels e-f) full nonlinear simulation with Athena++. The left and right rows show inner and outer disc, respectively. As a result of nonlinear effects (i.e. as compared to the top row showing the linear results), the wave steepens, shocks and subsequently decays in amplitude (while getting stretched azimuthally). In the inner disc, one can also see the emergence and nonlinear steepening of the secondary arm. }
\label{fig:lin-nonlin}
\end{figure*}

\begin{figure}
\centering
\includegraphics[width=0.49\textwidth]{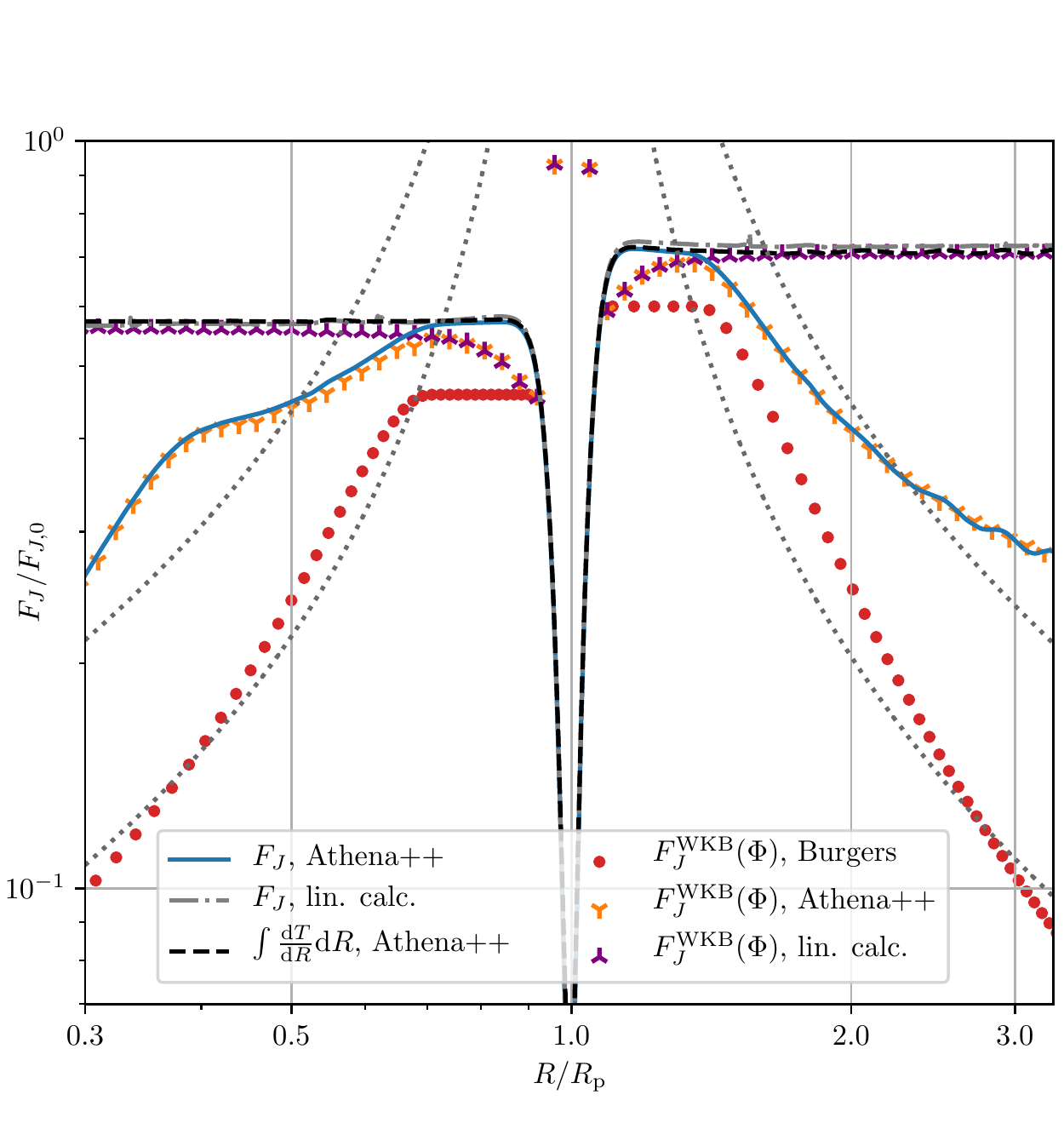}
\vspace*{-0.8cm}
\caption{A detailed comparison of different wave angular momentum metrics for $M_\mathrm{p} = 0.01 M_\mathrm{th}$. We compare the results for $F_J$ obtained using Eq. (\ref{eq:FJ}) (left column in the legend) as well as calculated using the WKB approximation via Eq. (\ref{eq:FJPhi}) ($F_J^\mathrm{WKB}$, right column in the legend). AMFs computed in the linear approximation (\S\ref{sect:linear}) are shown with grey dash-dotted and purple tripod curves; $F_J^\mathrm{WKB}$ obtained using the solutions to Burgers equation is shown with red dots; $F_J$ and $F_J^\mathrm{WKB}$ computed using Athena++ models are shown via blue solid and orange points; black dashed curve shows the integrated torque. All fluxes are rescaled by the characteristic amplitude $F_{J,0}$. Dotted grey lines indicate the scaling expected for an $N$-wave, $F_J \propto \tau^{-1/2}$, with two different normalizations.}
\label{fig:AMF001}
\end{figure}

Since the planet lies in the simulation domain we smooth the planet's potential to avoid the singularity at $\vect{r} = \vect{r}_\mathrm{p}$. In our fiducial setup we use a potential of the form \citep{Dong2011I}
\begin{align}
	\Phi_\mathrm{p}=\Phi_\mathrm{p}^{(4)} = -G M_\mathrm{p} \frac{d^2 + (3/2) r_\mathrm{s}^2}{\left(d^2 + r_\mathrm{s}^2 \right)^{3/2}},
	\label{eq:phi4}
\end{align}
where $d = \vert \vect{r} - \vect{r}_\mathrm{p} \vert$ is the distance from the planet and $r_\mathrm{s} = \epsilon H(R=R_\mathrm{p})$ is a smoothing length. This form of the potential has been used by \citet{Dong2011I,Dong2011II}, who have shown it to give better agreement with linear shearing-sheet calculations that do not employ any smoothing, than the typical Plummer-type potential used in literature, see Appendix \ref{sec:potentialorder} for more details. In cases where a different potential is used, we state it explicitly. We adopt $\epsilon = 0.6$ in this work, following \citet{Mueller2012}.

In order to avoid spurious shocks, we introduce the planetary gravitational potential over a timescale $t_\mathrm{ramp}$ that is typically set to 10 planet orbits, except for the highest resolution simulation in which we reduced it to 5 planet orbits (since it cannot be evolved for a very long time). Explicitly, the planet mass is introduced as
\begin{align}
	M_\mathrm{p}(t) = M_\mathrm{p}\begin{cases}
	 (1/2)\left[1-\cos\left(\pi t/t_\mathrm{ramp}\right) \right] & \mathrm{if} \,\, t < t_\mathrm{ramp},\\
	1 & \mathrm{else.}
	\end{cases}
\end{align}

We also employ an orbital advection scheme that is based on the FARGO algorithm \citep{FARGO2000}. Its implementation in Athena is described by \citet{Sorathia2012}. It has now also been implemented in Athena++ and is available in the latest public release (version 21.0). In our models orbital advection is performed at intermediate integration time-steps to keep the scheme second order accurate in time. In our version, orbital advection is done using the Keplerian background velocity $\vect{u}_\mathrm{K}(R) = \sqrt{GM_\star/R} \hat{e}_\phi$, which is only a function of $R$ and constant in time. This allows for a larger time-step, leading to a net speed-up of more than one order of magnitude for our typical setup. We find that using this method leads to decreased numerical diffusion as compared to models that do not employ this scheme. We have tested the method carefully, including tests with discs that do not host planets. As an additional check, we also performed several simulations using FARGO3D \citep{FARGO3D} and found good agreement between the results obtained with two different codes, see Appendix \ref{sec:fargo3d} for details.

\changed{
We vary the parameters of the problem in the following way: the planet mass is $\Mp / \Mth \in [0.01,0.1,0.25,0.5,1]$,
the disc aspect ratio at the planet location is $\hp \in [0.05,0.07,0.1]$ and the background disc surface density slope is $p \in [0,1.5]$.
}

The simulation domain extends over a radial range $0.2 \leq R/R_\mathrm{p} < 4.0$ with logarithmic spacing and uniformly extends over the full azimuthal angle $0 \leq \phi \leq 2\pi$.
We employ wave-damping zones \citep{dvB2007II} close to the radial boundaries in the zones $0.2 \leq R/R_\mathrm{p} \leq 0.28$ and $3.4 \leq R/R_\mathrm{p} \leq 4$ to avoid reflections. Only the radial velocity and density are damped towards their initial values, while the azimuthal velocity remains unchanged.

Probing the evolution of very small perturbations in the inviscid regime requires very high resolution to suppress numerical viscosity, which can lead to linear damping of the wave \citep[e.g.][]{Dong2011I,Dong2011II}. Regarding this issue, we perform an extensive resolution study to judge the convergence of our results. Resolutions ranged from the lowest $N_R \times N_\phi = 3448 \times 7200$ up to $N_R \times N_\phi = 27584 \times 57600$. This corresponds to 58 and 464 cells per disc scale-height at the planet radius.

We evolved the lowest resolution cases for about 1000 planetary orbits, while the highest resolution simulations were evolved for 20 orbits, due to their high computational cost.\footnote{These simulations used a total of 8192 threads, distributed over 32 nodes hosting Intel\textsuperscript{\textregistered} Xeon Phi\textsuperscript{TM}, running 256 threads each.} However, this is still long enough for the wave to reach a quasi-steady state
\changed{, as the longest sound crossing time across the domain (for $\hp = 0.05$) is less than 20 planet orbits.}

During the simulations we keep track of the fluxes computed by the Riemann solver, flux divergences and source terms in the conservation equations, 
as well as azimuthal averages of all conserved quantities and their time-derivatives on a time resolution given by the actual time-stepping of the code. We calculate the vorticity as a line-integral over cell edges using reconstructed values of velocities:
\begin{align}
	\omega_z = \int_C \left( \nabla \times \vect{u} \right) \, \de A = \oint_{\partial C} \vect{u} \cdot \,\de\vect{s}
\end{align}
and divide it by the cell-centred surface density to compute the vortensity in a cell. This method avoids calculating the velocity gradients in post-processing which can lead to diverging results close to discontinuities (shocks).


\subsection{Linear calculations}
\label{sect:linear}


To provide a benchmark for Athena++ to meet in the low planet mass limit ($M_\mathrm{p} \ll M_\mathrm{th}$), we use the numerical method of \citet{Miranda2019I} to calculate the global structure of linear perturbations in globally isothermal discs. This calculation, relying on solving the linearized fluid equations in fully global setup, allows us to confirm the precision of the implemented methods in fully nonlinear simulations and to demonstrate the transition into weakly nonlinear behaviour. For consistency, we employ the same potential $\Phi_\mathrm{p}^{(4)}$ in these calculations (see also Appendix \ref{sec:potentialorder}).


\subsection{Solutions of Burgers equation}
\label{sect:Burgers}


Our verification of the weakly nonlinear theory described in \S\ref{sect:wave-nonlin} relies on solving the inviscid Burgers equation (\ref{eq:Burger}) in order to directly obtain wave profiles and to compare their evolution with our simulations results. We numerically solve Eq. (\ref{eq:Burger}) adopting a finite-volume scheme with Engquist-Osher flux-splitting and a second-order Runge-Kutta integrator for stepping in $\tau$. As initial wave profiles, we use data from our nonlinear simulations, close to the planet at $\tau = \tau_0$, where excitation of the wake is expected to be almost complete. This corresponds to $R/R_\mathrm{p} \simeq 0.936$ and $R/R_\mathrm{p} \simeq 1.068$ (in the inner and outer disc) for $h_\mathrm{p} = 0.05$.


\section{Evolution of the Planet-driven waves as a function of $M_\mathrm{p}$}
\label{sec:res}


In this section we compare different descriptions of the planet-driven density wave evolution for two different planet masses assuming the fiducial disc with a density slope $p = 3/2$ and aspect ratio at planet location $h_\mathrm{p} = 0.05$ (dependence on disc parameters is explored in \S\ref{sec:resdisc}). Unless specifically mentioned, e.g. when discussing long-term behaviour, results are shown after 20 planetary orbits, when the initial distribution of $\Sigma$ in the disc is not yet strongly perturbed.

\begin{figure*}
\centering
\includegraphics[width=\textwidth]{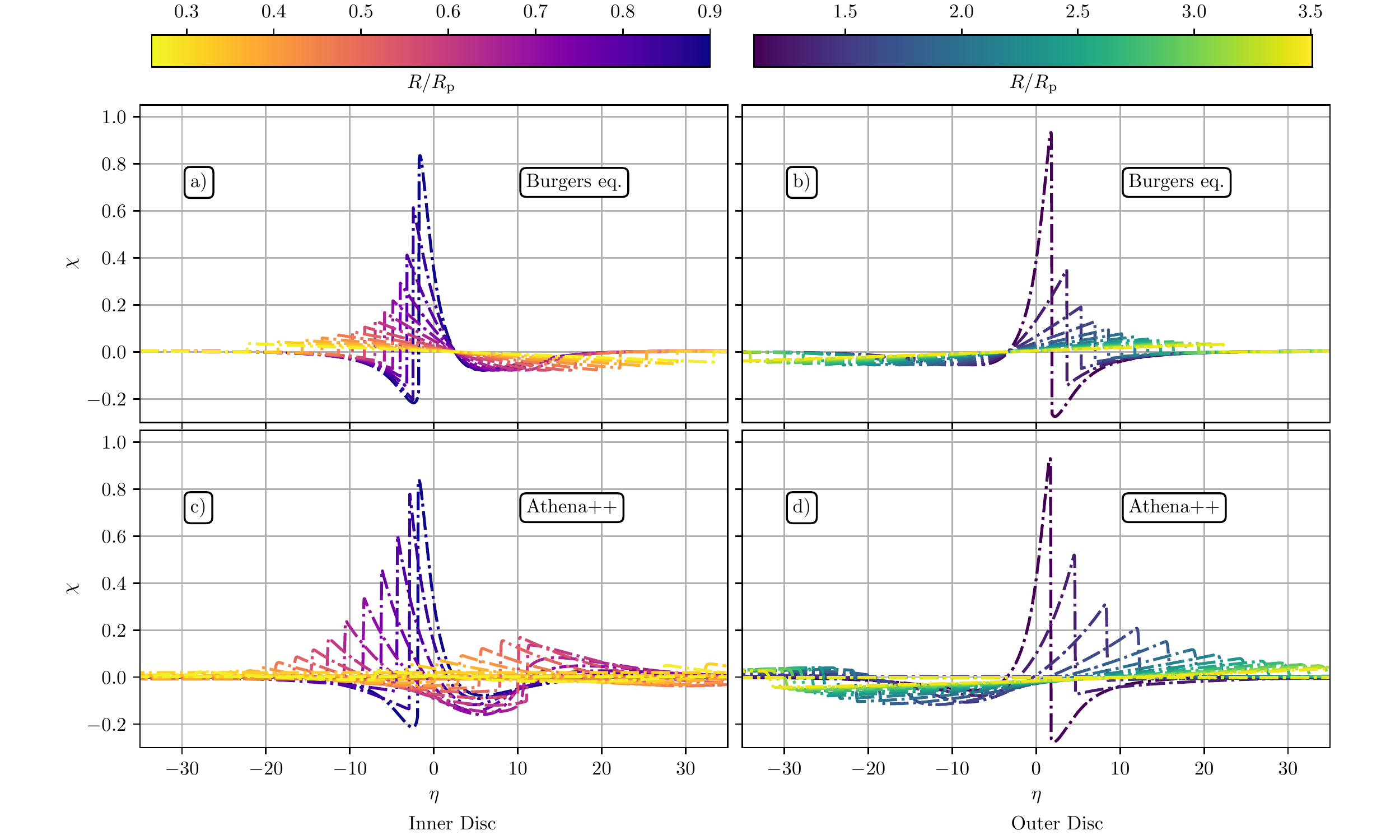}
\vspace*{-0.3cm}
\caption{Same as middle and bottom rows of Fig. \ref{fig:lin-nonlin} but for an intermediate-mass planet, $M_\mathrm{p} = 0.25 M_\mathrm{th}$. Note a faster nonlinear evolution of the wave profile.}
\label{fig:025nonlin-burger}
\end{figure*}

\begin{figure}
\centering
\includegraphics[width=0.49\textwidth]{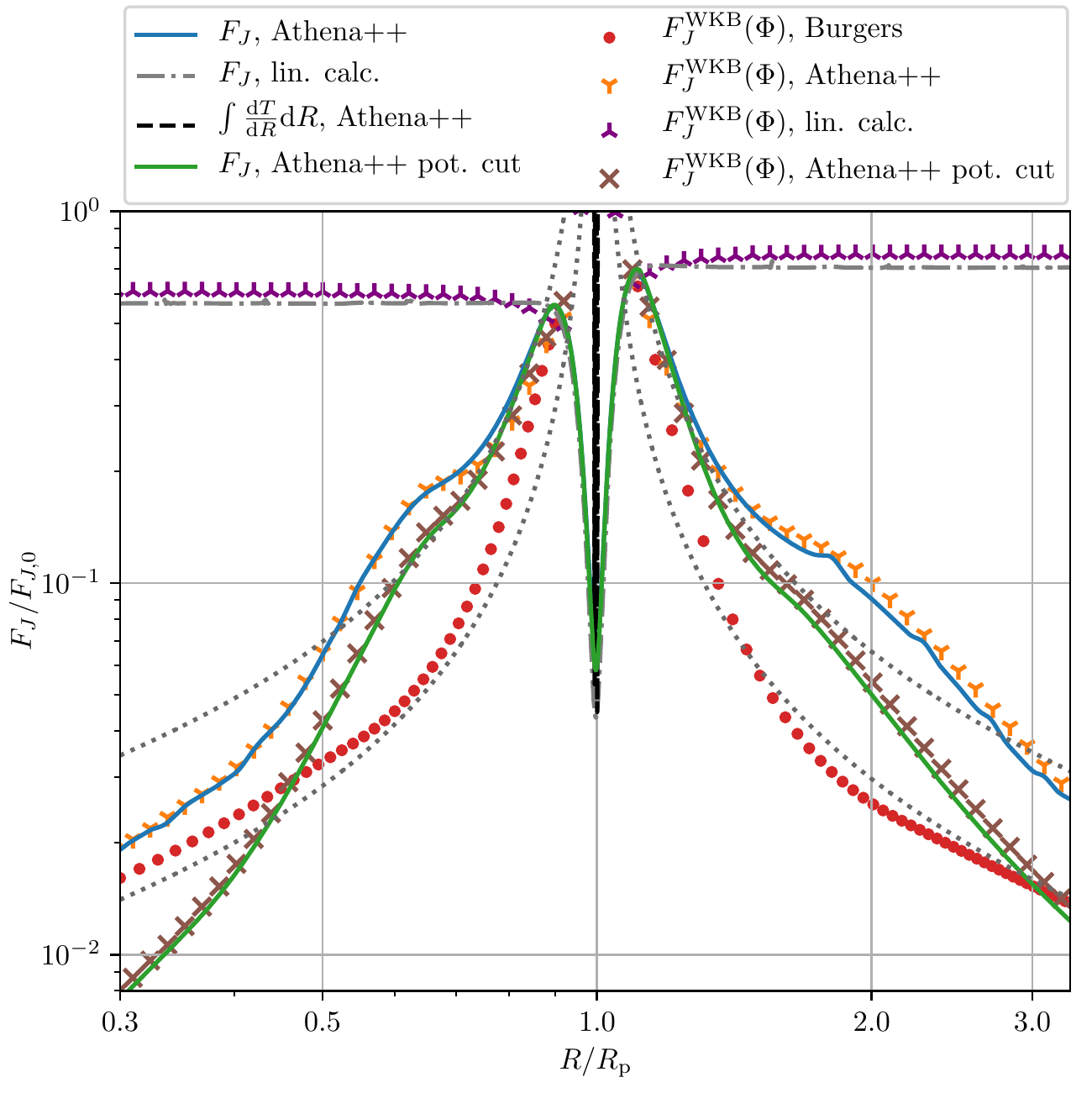}
\vspace*{-0.7cm}
\caption{Same as Fig. \ref{fig:AMF001} but for $M_\mathrm{p} = 0.25 M_\mathrm{th}$. Results from the linear calculation are identical to those presented in \S\ref{eq:low-mass} and are shown just for reference. Green curve and brown crosses show the results of a $M_\mathrm{p} = 0.25 M_\mathrm{th}$ run with a spatially truncated potential (see Eq. (\ref{eq:phi-trunc})), discussed in detail in \S\ref{ssec:globwave}. Note that AMFs show no extended plateau after wave excitation, as for this $M_\mathrm{p}$ the shock forms at a distance
\changed{
$l_\mathrm{sh} \simeq 0.075 R_\mathrm{p}$
}
from the planet. The higher amplitude wave leads to a stronger shock, resulting in faster wave decay. See text for details.}
\label{fig:AMF025}
\end{figure}


\subsection{The low-mass case, $M_\mathrm{p} = 0.01 M_\mathrm{th}$}
\label{eq:low-mass}


In order to verify our implementation of the planetary potential and as an additional check of the orbital advection algorithm in Athena++, we compare the results of a simulation featuring a low-mass planet $M_\mathrm{p} = 0.01 M_\mathrm{th}$ to the solutions of the linear problem obtained from the method described in \citet{Miranda2019I}, see \S\ref{sect:linear}. Since $l_\mathrm{sh}\approx 5.4H_\mathrm{p}$ for this $M_\mathrm{p}$, the regions of wave excitation (within $(1-2)H_\mathrm{p}$ from the planet) and weakly nonlinear propagation should be spatially well-separated. This particular run features the highest resolution we explored ($N_R \times N_\phi = 27584 \times 57600$), which is needed to correctly capture the weak nonlinearity of the low-amplitude wave. Lower resolutions show signs of purely numerical wave decay before the shock is formed (see also \S\ref{sec:resvort}). 

We start by displaying in Fig. \ref{fig:lin-nonlin} azimuthal (or $\eta$) slices of the dimensionless wave perturbation $\chi$ defined by equation (\ref{eq:chi_g}), at different radii $R$. We compare results of linear calculations (top row), solutions of the Burgers equation (middle row) and the fully nonlinear simulation results (bottom row). The left (right) panel show the inner (outer) wake. As has been pointed out by \citet{Bae2018} and \citet{Miranda2019I}, a secondary arm naturally appears in the linear calculation at $R\lesssim 0.5R_\mathrm{p}$, without the need for nonlinearity, see panel (a). There we also see that the peak of the primary arm decreases in amplitude (and slightly shifts horizontally leftward), as a result of its AMF being transferred to the secondary arm. In the outer disc, the linear calculation (panel b) predicts a single peak that stays around $\eta=0$ with fixed amplitude and no additional arms are formed. Some transfer of AMF occurs only between the leading and trailing troughs surrounding the peak. Due to the linear nature of this calculation, wave breaking does not occur.

Moving to solutions of the Burgers equation (panels c \& d), we see that the inner and outer disc evolve very similarly, as expected from the form of the equation. Since this equation does not account for the linear wave evolution, its solutions do not capture the formation of the secondary arm in the inner disc and transfer of flux between the troughs in the outer disc. The nonlinear nature of this approximation leads to steady wave steepening that eventually leads to the formation of a shock (a sharp jump in the wave profile). After the shock has formed, the resulting dissipation gradually reduces the wave amplitude while the wave profile broadens.

Comparing the top and bottom rows, we find excellent agreement in both the shape and amplitude of the excited wake close to the planet, confirming that our implementation in Athena++ reproduces the linear prediction for $M_\mathrm{p} \ll M_\mathrm{th}$. But as the distance from the planet increases, the wave profile in a simulation (panels e \& f) gets additionally distorted by nonlinear effects, leading to a horizontal shift of its peak to more positive (negative) values of $\eta$ as compared to the linear prediction in the outer (inner) disc. After a shock is formed, the wave starts to decay due to dissipation. Note that far from the planet $(R \gtrsim 2.4)$ and $(R \lesssim 0.5)$, the wakes in panels (e) \& (f) do not display such steep jumps as the solutions of Burgers equation, most likely because of wave dispersion (and some numerical diffusion). This comparison clearly shows that the full calculation (bottom row) naturally combines linear and nonlinear effects, with both playing important roles.

To further understand the different approximations used in this work, in Fig. \ref{fig:AMF001} we show the rescaled angular momentum flux $F_J/F_{J,0}$ (see Eq. (\ref{eq:FJ0})) computed in different ways, as a function of radius $R$. For the Athena++ simulations and the linear calculation, we compare results for $F_J^\mathrm{WKB}$ in the WKB approximation (i.e. with $F_J$ obtained from $\Phi$, see Eq. \ref{eq:FJPhi}) to $F_J$ obtained
directly from the full solutions for density and velocity perturbations, see equation (\ref{eq:FJ}). In the case of Burgers equation that has been reduced to the variable $\chi$, we show only the former.\footnote{The divergent behaviour of $\Phi$ around $R=R_\mathrm{p}$ is due to the fact that the atmosphere of the planet leads to a finite surface density perturbation there, while $g(R)$ in the definition (\ref{eq:chi_g}) of $\chi$ diverges as $R\to R_\mathrm{p}$.}

In the linear calculation (grey dash-dotted) $F_J$ first increases with $|R-R_\mathrm{p}|$ close to the planet, before reaching a plateau a few scale-heights away from the planet. At the same time, the linear $F_J^\mathrm{WKB}$ disagrees with the corresponding $F_J$ close to the planet. This disagreement is expected, since the wake becomes tightly wound (and WKB approximation becomes accurate) only after propagating a certain distance away from the planet. Beyond $\simeq 0.3 R_\mathrm{p}$ from the planet, both ways of computing AMF agree to within a few percent. Note the asymmetry in the saturated $F_J$ levels between the inner and outer discs, that is ultimately responsible for planet migration. 

The $F_J^\mathrm{WKB}$ computed using the solution of Burgers equation (red dots) shows an initially constant flux (which is below the linear prediction since the initial condition for the Burgers evolution was set close to the planet, where the linear $F_J$ has not yet fully accumulated), until the shock is formed $\simeq 0.3 R_\mathrm{p}$ away from the planet, in good agreement with the predicted shocking distance $l_\mathrm{sh} \simeq 0.27R_\mathrm{p}$. Beyond that point, the Burgers AMF decays and asymptotically follows the $N$-wave scaling, $F_J \propto \Phi \propto \tau^{-1/2}$ (grey dotted curves with different normalization), although convergence to the asymptotic scaling is slow, as expected from \citetalias{Goodman2001} and the fact that $\tau$ is small for low $M_\mathrm{p}$, see equation (\ref{eq:tau}). 

Finally, for the $F_J$ derived from our simulations (blue solid curve) we find excellent agreement with the linear $F_J$ close to the planet, for $|R-R_\mathrm{p}|\lesssim l_\mathrm{sh}$ --- the location and amplitude of the peak $F_J$ match precisely. This means that $F_J$ is accurately conserved after excitation, indicating negligible numerical dissipation. To provide yet another check, we plot the integrated torque density from the simulation with the black dotted curve, which agrees perfectly with $F_J$ close to the planet (and with the linear $F_J$ far from it).

At a distance that agrees with the theoretical prediction $l_\mathrm{sh}$ and results from Burgers equation, the wave in our simulation shocks and starts to decrease. Its $F_J$ drops below the linear $F_J$ and the integrated torque density. At the same time, both in the inner and outer disc, $F_J$ decays notably slower than what Burgers equation predicts. The slow decay in the inner disc between $0.4 \lesssim R/R_\mathrm{p} \lesssim 0.6$ can be easily recognized as being caused by the formation of the
secondary arm, see Fig. \ref{fig:lin-nonlin}. But even in the outer disc, where this effect is absent, the $F_J$ at the damping boundary in the full simulation is still 2-3 times higher than $F_J^\mathrm{WKB}$ resulting from the Burgers evolution.


\subsection{An intermediate mass planet, $M_\mathrm{p} = 0.25 M_\mathrm{th}$}


We now examine the case of a more massive planet $M_\mathrm{p} = 0.25 M_\mathrm{th}$, keeping the same (fiducial) disc parameters.

In Fig. \ref{fig:025nonlin-burger}, we again plot azimuthal wake profiles, this time omitting the linear calculation, which is no longer relevant for this $M_\mathrm{p}$ because of the increased importance of nonlinear effects. As compared to the low-$M_\mathrm{p}$ case, we see that the shock forms closer to the planet and the wave evolves faster, e.g. the wake decays and broadens azimuthally more rapidly. This is because equation (\ref{eq:tau}) yields larger values of the time-like coordinate $\tau$ for higher $M_\mathrm{p}$, everything else being equal. As expected, solutions of the Burgers equation again differ from the full simulation results by not capturing the formation of the secondary spiral arm in the inner disc, see panels (a) and (c). Also, similar to the low-$M_\mathrm{p}$ case shown in \S\ref{eq:low-mass}, Burgers framework (top row) predicts a stronger wave decay and a slower advance of the shock front compared to the simulation (bottom row) at intermediate distances from the planet ($R/R_\mathrm{p} \lesssim 0.8$ and $1.3 \lesssim R/R_\mathrm{p}$). Nevertheless, the evolution of the wake shape in an $M_\mathrm{p} = 0.25 M_\mathrm{th}$ simulation bears more resemblance to the Burgers equation solutions than in the $M_\mathrm{p} = 0.01 M_\mathrm{th}$ case, because of the increased importance of wave nonlinearity.

These conclusions are corroborated by the behaviour of the angular momentum fluxes in $M_\mathrm{p} = 0.25 M_\mathrm{th}$ case, see Fig. \ref{fig:AMF025}. In a simulation, shortly after reaching the maximum value that agrees with the integrated torque density, $F_J$ begins to decay as a result of wave shocking at
\changed{
$l_\mathrm{sh} \simeq 0.075 R_\mathrm{p}$
}
away from the planet. The $F_J$ damping is faster for this higher $M_\mathrm{p}$, and the AMF decreases by more than a factor of 10 when the wave reaches the damping zones. In the inner disc this decay is again delayed by the formation of the secondary arm, but overall, both in the inner and outer discs, $F_J$ is not too deviant from the $\tau^{-1/2}$ scaling expected from the Burgers framework, which is reasonably well followed by the Burgers $F_J^\mathrm{WKB}$ (red dots). Even though $F_J$ in a simulation still decays with the distance slower than the Burgers $F_J^\mathrm{WKB}$, the disagreement between the two is reduced for higher $M_\mathrm{p}$.


\section{Characteristics of the planet-driven spiral shocks}
\label{sec:shock}


In agreement with the local version of the nonlinear evolution framework (\ref{eq:tau})-(\ref{eq:Burger}), \citet{Dong2011II} found that shocks forming in their simulations as a result of nonlinear evolution of planet-driven density waves possess certain universal characteristics. We now examine whether the same is true in our global setup. 


\subsection{Evolution of the shock strength}
\label{sec:deltachi}


\begin{figure}
\centering
\includegraphics[width=0.49\textwidth]{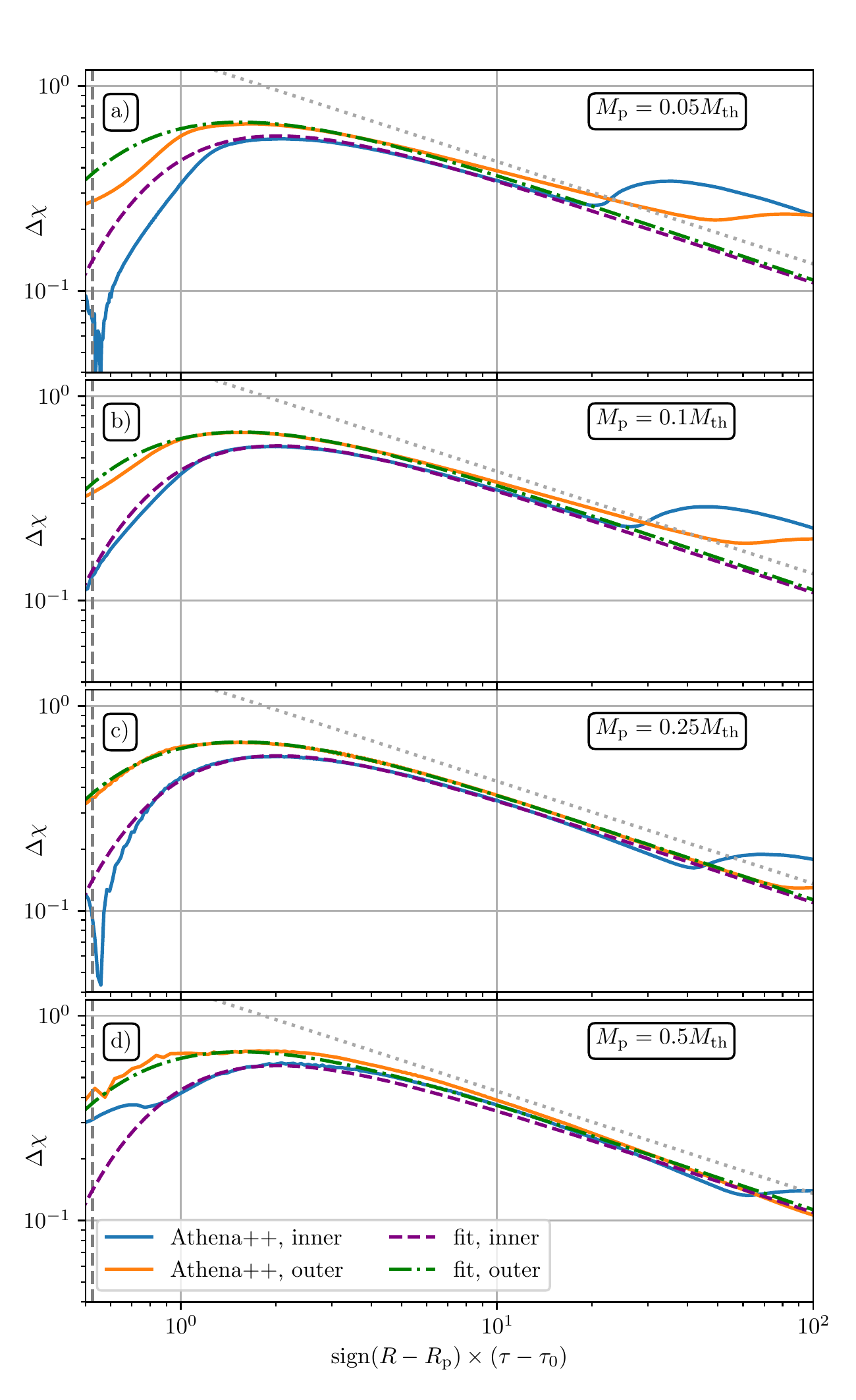}
\vspace*{-0.7cm}
\caption{Shock strength $\Delta \chi$ (jump of $\chi$ across the shock) obtained from our simulations using a method described in \S\ref{sec:deltachi} as a function of the coordinate $\tau$, for different planet masses $M_\mathrm{p}$ and the fiducial disc parameters $p=3/2, h =0.05$. The blue and orange curves show results in the outer and inner disc, respectively. The same value of $\tau$ corresponds to different radial separations from the planet, depending on planet mass (see Eq. (\ref{eq:tau})). As expected from the Burgers equation framework, there is a strong similarity in $\Delta\chi(\tau)$ behavior across all masses. The green dotted and purple dashed lines show a fit (in the outer and inner disc, correspondingly) using a smooth broken two-component power-law, see Eqs. (\ref{eq:fit}). Grey dash-dotted lines show the scaling $\Delta \chi \propto \tau^{-1/2}$ expected for large $\tau$.}
\label{fig:dchi-t}
\end{figure}

The main characteristic of a shock wave is its strength $\epsilon \equiv \Delta \Sigma / \Sigma$, where $\Delta\Sigma$ is the surface density jump (in 2D) across the shock front. Measuring $\epsilon$ in isothermal simulations can be surprisingly difficult (especially for weak shocks), which has been noted before. This issue does not arise for non-isothermal shocks, for which the entropy jump can be used to effectively estimate the shock strength \citep{Arza2018}. Previous studies have often used the peak value of the surface density perturbation to determine the shock strength \citep[e.g.][]{Ziampras2020I}. This approximation would be reasonable for a density wave that has already evolved into an $N$-wave. However, it will fail for the waves that are just starting to develop shocked segments.  

At the next level of sophistication one can compare the minimum and maximum values of the density perturbation $\delta \Sigma/\Sigma_0$ inside some azimuthal interval $\delta \phi$ around the discontinuity. We found that the results of this method depend strongly on the choice of $\delta \phi$, which can lead to inconsistent results.

For these reasons, we use an alternative method to obtain $\epsilon$ from our simulation results, which exploits a well known fact that the energy and angular momentum losses of a density wave are connected (\citealt{Lynden1972}; \citetalias{Goodman2001}). Using this argument, one can show \citep{Rafikov2016} that the {\it angular momentum dissipation} rate (per unit radial distance) of a spiral shock with $m$-fold azimuthal symmetry and pattern frequency $\Omega_\mathrm{p}$ is given by\footnote{We have replaced the pre-shock density $\Sigma_\mathrm{pre}$ as in \citet{Arza2018} with the azimuthally-averaged density $\langle \Sigma \rangle_\phi$ as in \citet{Rafikov2016}.}
 \begin{align}
	\left.\pd{F_J}{R}\right\vert_{\mathrm{diss}} = \operatorname{sign} \left( \Omega_\mathrm{p} - \Omega(R) \right) m R \langle \Sigma \rangle_\phi \cs^2 \Psi_Q(\epsilon),
	\label{eq:wavedamp}
\end{align}
where for an isothermal EoS \citep{Belyaev2013,Rafikov2016}
\begin{align}
    			\Psi_Q(\epsilon) = \frac{\epsilon \left( 2 + \epsilon \right) - 2\left(1 + \epsilon \right) \ln \left( 1 + \epsilon \right)}{2\left( 1 + \epsilon \right)}.
    			\label{eq:PsiQ}
\end{align}

In the case of a planet-driven shock $\partial_R F_J\vert_{\mathrm{diss}}$ is in general not equal to the radial derivative of the AMF, $\partial_R F_J$. This is because the latter is also affected by the torque deposition due to the planetary potential, $\mathrm{d}T/\mathrm{d}R$, so that
\begin{align}
	\label{eq:dfdr}
	\pd{F_J}{R} = \left.\pd{F_J}{R}\right\vert_{\mathrm{diss}} + \td{T}{R}.
\end{align}

We use this relation in our simulations to find the angular momentum deposition rate due to the shock $\partial_R F_J\vert_{\mathrm{diss}}$ as the difference between the radial angular momentum flux divergence and the torque density (Eq. (\ref{eq:dTdR})) directly computed from the simulation data (i.e. we use divergences of the Riemann fluxes and the source term monitored in Athena++ on the run). At radii where there is only one shock (i.e. that of the primary arm), the knowledge of $\partial_R F_J\vert_{\mathrm{diss}}$ then allows us to deduce the shock-strength $\epsilon$ by solving Eqs.
(\ref{eq:wavedamp})-(\ref{eq:PsiQ}) for $\epsilon$, assuming $m=1$ and using a Newton-Raphson root finder. We note that this method requires high resolution simulations to ensure that linear damping due to numerical viscosity does not significantly affect $\partial_R F_J$, which would otherwise bias our estimate of $\epsilon$.

We illustrate the application of this procedure to our simulation results in Fig. \ref{fig:dchi-t}, where we display the shock strength expressed in terms of $\Delta \chi$ --- the jump of $\chi$ across the shock, which can be related to $\epsilon$ via Eq. (\ref{eq:chi_g}). We show $\Delta \chi$ as a function of the coordinate $\tau$ for several planet masses (increasing from top to bottom) and fiducial disc parameters, for both the inner (blue) and outer (orange) disc. This figure clearly reveals a rather universal evolution of the shock strength: around $\tau = \tau_\mathrm{sh}$ defined by the Eq. (\ref{eq:taush}) a shock appears, $\Delta \chi$ becomes non-zero and rapidly reaches its maximum value, after which it gradually falls off, showing a scaling close to that expected from Burgers equation, $\Delta \chi \propto \tau^{-1/2}$. For all masses we find that the outer wake produces a stronger shock, consistent with higher wave amplitudes at excitation for the fiducial disc parameters. Comparing $\Delta \chi$ obtained by this method to the jumps in azimuthal profiles of $\chi$ in Figs. \ref{fig:lin-nonlin}, \ref{fig:025nonlin-burger}, we find good agreement.

\begin{table}
\caption{Parameters of the fit to $\Delta \chi (\tau)$ (see Eq. (\ref{eq:fit})).}
\label{tab:fitpars}
  \begin{tabular}{lccccc}
    \hline
    & $A$ & $\tilde{\tau}_\mathrm{b}$ & $\alpha_1$ & $\alpha_2$ & $\Delta$ \\
    \hline
    inner disc & 2.07 &  0.300 & -10.84 &  0.505 & 0.623\\
    outer disc & 3.11 & 0.181 & -8.63 &  0.525 & 0.766\\
    \hline
  \end{tabular}
\end{table}

The previously noted approximate universality of the shock strength behavior motivates us to provide a fitting formula that would uniformly approximate $\Delta \chi(\tau)$ behavior in Fig. \ref{fig:dchi-t} for different $M_\mathrm{p}$. To this effect we converged on a smoothly broken two-component power law for $\Delta \chi$ as a function of $\tau$ with $\tilde{\tau} \equiv \tau - \tau_0$ (see Eq. \ref{eq:tau0} for $\tau_0$):
\begin{align}
	\Delta \chi(\tau) = A \left(\frac{\tilde{\tau}}{\tilde{\tau}_\mathrm{b}}\right)^{-\alpha_1} \left\lbrace \left[ 1 + \left(\frac{\tilde{\tau}}{\tilde{\tau}_\mathrm{b}}\right)^{1/\Delta}\right]\right\rbrace^{(\alpha_1 - \alpha_2)\Delta},
	\label{eq:fit}
\end{align}
where $A$ is an amplitude and $\tilde{\tau}_\mathrm{b}$ marks the breaking point between the two asymptotic slopes $\alpha_{1}$ and $\alpha_{2}$. The parameter $\Delta$ determines the width of this transition. Our derived set of the fit parameters is shown in Table \ref{tab:fitpars}. It best approximates the results of the intermediate mass $M_\mathrm{p} = 0.25 M_\mathrm{th}$ simulation, see Fig. \ref{fig:dchi-t}c. Despite certain deviations of this fit from the data in the very low $M_\mathrm{p}\ll M_\mathrm{th}$ and $M_\mathrm{p}\sim M_\mathrm{th}$ regimes, it still covers a broad range of $M_\mathrm{p}$ values reasonably well.

Note that for large values of $\tau$ in the inner disc, the shock strength inferred from the simulations displays a second peak. It is associated with the emergence of the secondary spiral arm and its evolution into a shock. Since the appearance of a secondary arm is a linear effect \citep{Miranda2019I}, taking place at a fixed distance from the planet, the shift of the second peak towards higher $\tau$ for larger $M_\mathrm{p}$ can be well understood using Eq. (\ref{eq:tau}). Once the secondary shock has formed, our fit (\ref{eq:fit}) no longer works in the inner disc.

In the outer disc, the rising $\Delta \chi$ that is seen for large $\tau$ at low planet masses ($0.05 - 0.25 \Mth$) is simply due to the fact that the wave reaches
the damping zones, making our estimate of the shock strength invalid.


\subsection{Location of the shock front}
\label{sec:phishock}


Another important characteristic of a planet-driven density wave is its global shape in the $R-\phi$ coordinates. Understanding the factors determining the shape of a density wave is crucial for interpreting observations of the spiral arms in protoplanetary discs \citep[e.g.][]{Zhu2015,Bae2018}. Also, as we will show in \S\ref{sec:resvort}, the geometry (curvature) of the spiral shock has direct effect on the generation of vortensity. Here we outline theoretical expectations for the shape of the spiral shock and compare them to simulations.

In linear theory the wave shape (defined as the $R-\phi$ curve along which $\eta=\const$) should be given by $\phi=\phi_\mathrm{lin}(R)$, see Eq. (\ref{eq:phi_lin}). However, nonlinear effects cause broadening of the wake, which steadily displaces wake maximum in the azimuthal direction compared to the linear prediction, see Fig. \ref{fig:025nonlin-burger} for a clear illustration of this phenomenon (note that $\eta\propto \phi-\phi_\mathrm{lin}$). 

Weakly nonlinear wave evolution theory (\citealt{LL}; \citetalias{Goodman2001}) predicts that, in the N-wave regime, the broadening of the wake in the $\eta$ coordinate (as well as the $\eta$-displacement of its peak) behaves as $\Delta\eta\propto (\tau-\tau_0)^{1/2}$, where $\tau_0$ (Eq. (\ref{eq:tau0})) corresponds to the point where wave excitation is mostly complete \citepalias{Goodman2001}. Then Eq. (\ref{eq:eta}) suggests that the azimuthal deviation $\Delta \phi_\mathrm{sh} \equiv \phi_\mathrm{sh} - \phi_\mathrm{lin}$ of the shock position from the linear prediction $\phi_\mathrm{lin}$ should scale as 
\begin{align}
	\Delta \phi_\mathrm{sh} \propto \operatorname{sign}(R - R_\mathrm{p})~ h_\mathrm{p}~(\tau-\tau_0)^{1/2}.
	\label{eq:deltaphi}
\end{align}
From this one would expect the position of the shock front corrected for the nonlinear effects to be given by 
\begin{align}
	\phi_\mathrm{sh}^\mathrm{nonlin}(R) =  \phi_\mathrm{lin}(R) + \operatorname{sign}(R - R_\mathrm{p}) ~\Delta \phi_0   ~h_\mathrm{p}~(\tau-\tau_0)^{1/2}.
	\label{eq:phinonlin}
\end{align}
where $\Delta \phi_0$ is the only fitting constant, which is fixed using our simulations as shown below. A similar approach has been used in  \citet{Zhu2015}, who studied spiral shocks for higher $\Mp\gtrsim \Mth$. 

\begin{figure}
\centering
\includegraphics[width=0.49\textwidth]{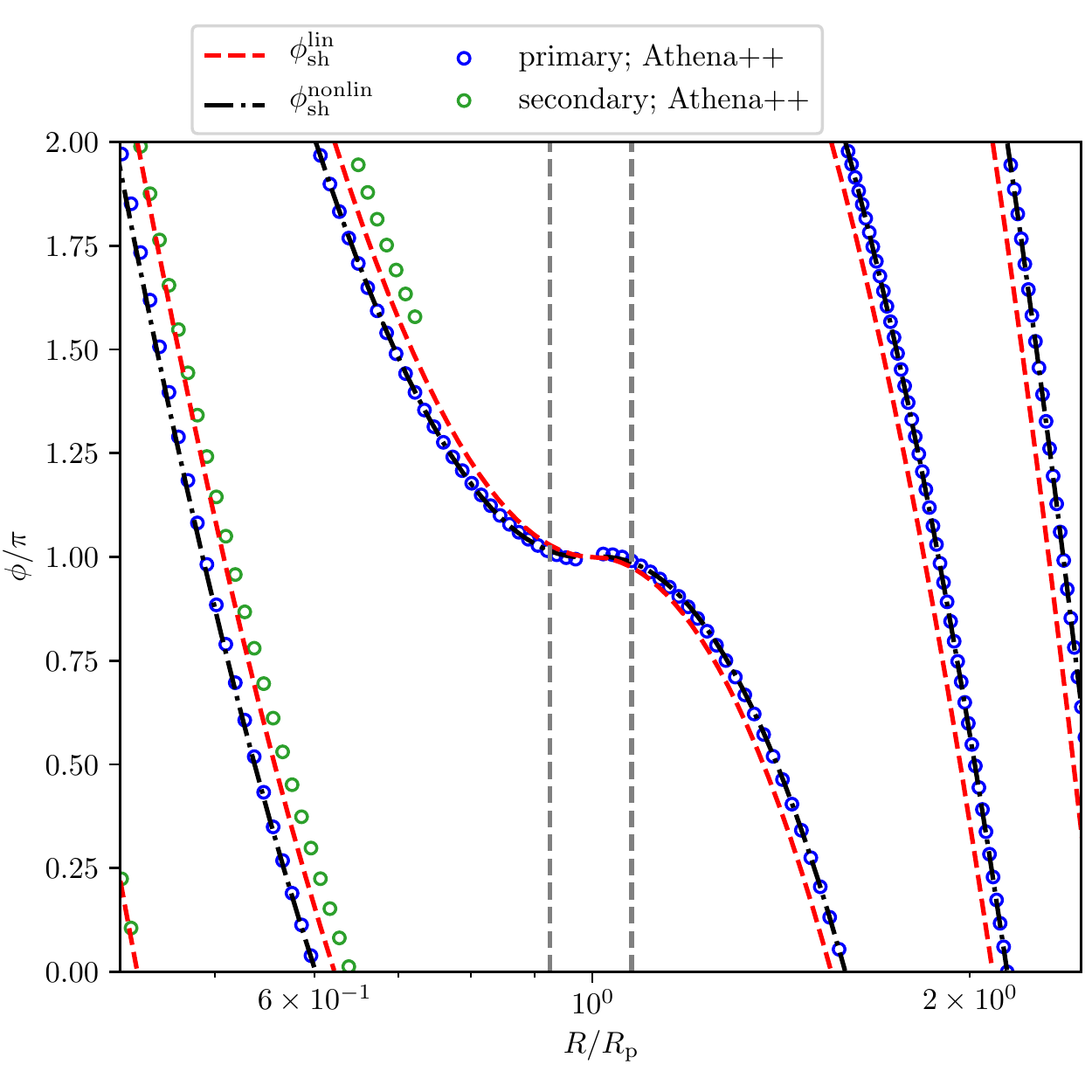}
\vspace*{-0.8cm}
\caption{Global pattern of the primary wave front, defined as the steepest part of the azimuthal density profile for the fiducial disc and planet with $M_\mathrm{p} = 0.25 M_\mathrm{th}$. Blue and green dots mark locations of the primary and secondary arm from the simulation, respectively. Red dashed curve follows the theoretical linear prediction (Eq. (\ref{eq:philin})) and black dash-dotted curve is a theoretical fit given by the Eq. (\ref{eq:phinonlin})-(\ref{eq:phi-pars}). The radial separation $l_\mathrm{sh}$ at which the shock forms is marked by vertical dashed lines. See \S\ref{sec:phishock} for details.  }
\label{fig:wake-location}
\end{figure}

\begin{figure}
\centering
\includegraphics[width=0.49\textwidth]{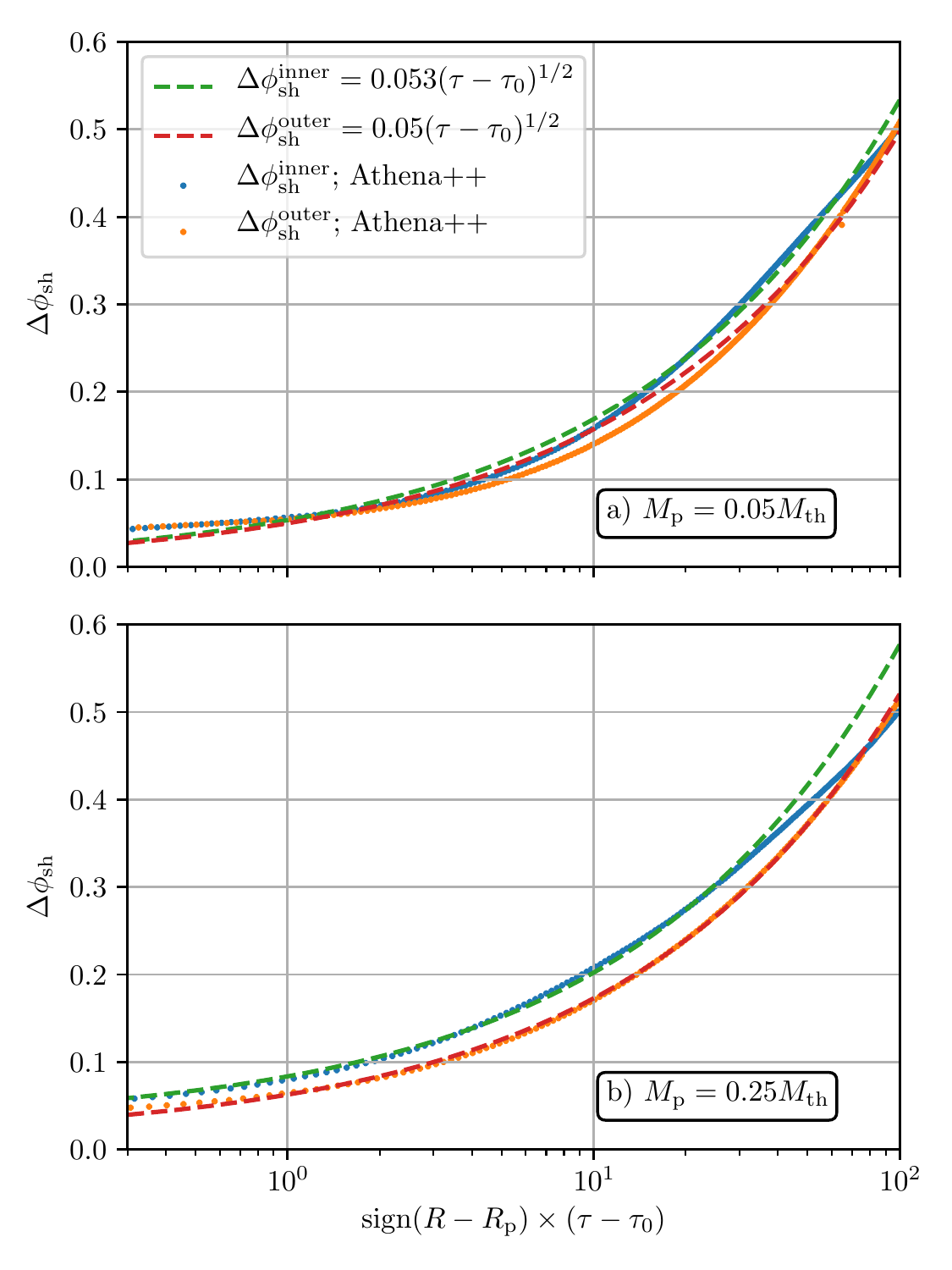}
\vspace*{-0.8cm}
\caption{Deviation of the azimuthal shock location from linear theory prediction $\Delta \phi = \phi_\mathrm{sh} - \phi_\mathrm{lin}$. Simulations results for fiducial disc parameters with $M_\mathrm{p} = 0.05 M_\mathrm{th}$ (top) and $M_\mathrm{p} = 0.25 M_\mathrm{th}$ (bottom) are shown. Blue and orange dots mark simulation results for the outer and inner disc, respectively. A fit for Eq. (\ref{eq:deltaphi}) is given by the red and green dashed lines, with a proportionality constant indicated in the legend. 
}
\label{fig:phi-deviation}
\end{figure}

In our runs the shock front is found by searching for a maximum value of $\partial_\phi \ln \Sigma$ at a fixed $R$, a method which has been used before by \citet{Arza2018} to locate the wave front as the steepest part of its profile. In the inner disc, care has to be taken at radii where the secondary arm forms and eventually creates a shock. As we have seen in previous sections, both linear and nonlinear effects can lead to a shift of the wave front, contributing to $\Delta\phi_\mathrm{sh}$ in the inner disc. Recall that even for the global linear calculation, the front of the primary wake is not centred at $\eta = 0$ (see Fig. \ref{fig:lin-nonlin}).

In Fig. \ref{fig:wake-location} we display the global wave pattern in a simulation of the fiducial disc model with $M_\mathrm{p} = 0.25 M_\mathrm{th}$, and compare it with both linear (Eq. [\ref{eq:phi_lin}]) and nonlinear (Eq. [\ref{eq:phinonlin}]) predictions. Around excitation, the linear wake prediction $\phi_\mathrm{lin}$ (centered on $\eta=0$) is already slightly offset from the wave front (see Fig. \ref{fig:lin-nonlin}a,b). As the wave propagates away from the planet, the disagreement in $\phi$ between the $\phi_\mathrm{lin}$ and the actual position of the shock increases, mainly due to the aforementioned nonlinear effects. Additionally, in the inner disc, the secondary arm forms behind the primary at $R \simeq 0.6 R_\mathrm{p}$, eventually passing through $\phi_\mathrm{lin}$. 

In Fig. \ref{fig:phi-deviation} we plot the azimuthal deviation from the linear prediction $\Delta \phi_\mathrm{sh}$ found in simulations as a function of $\tau$, for two different planet masses, $M_\mathrm{p} = 0.05 M_\mathrm{th}$ and $0.25 M_\mathrm{th}$. One can see that $\Delta \phi_\mathrm{sh}$ steadily increases with $\tau$ as the nonlinear effects accumulate in the course of wave propagation. We fit these numerical results with a behavior in the form (\ref{eq:deltaphi}), determining the values of the coefficients in front of the scaling as indicated in the legend of Fig. \ref{fig:phi-deviation}a (which uses $h_\mathrm{p}=0.05$). We find that in a fiducial disc the value of $\Delta \phi_0$ fitted to the numerical data varies by only a few percent as $M_\mathrm{p}$ changes from $0.05 M_\mathrm{th}$ to 0.5 $M_\mathrm{th}$. The dependence of $\Delta \phi_0$ on $h_\mathrm{p}$ or surface density slope is studied in \S\ref{sec:resdisc}, and shown to be negligible. This is expected as any dependence on these variables  and $M_\mathrm{p}$ is already absorbed in the Eq. (\ref{eq:phinonlin}) and the definition of $\tau$, see Eq. (\ref{eq:tau}).

The differences between the inner and outer discs are rather small, even despite the emergence of the secondary arm in the inner disc. Thus, for simplicity, in the following we use 
\begin{align}
	\Delta \phi_0 \approx 1,
	\label{eq:phi-pars}
\end{align}
which works well both in the inner and outer disc.

The nonlinear prediction (\ref{eq:phinonlin}) with thus fixed $\Delta \phi_0$ is shown as the blue dot-dashed curve in Fig. \ref{fig:wake-location}. One can see that it follows the location of the primary arm very well over a significant radial range both in the inner and outer discs. The prescription (\ref{eq:phinonlin})-(\ref{eq:phi-pars}) will be extensively used in the next section.


\section{Vortensity evolution due to planet-driven shocks}
\label{sec:resvort}


The vortensity (or potential vorticity) of a flow $\vect{\zeta}$ is given by the ratio of vorticity $\vect{\omega} = \nabla \times \vect{u}$, where $\vect{u}$ is the fluid velocity, and density $\rho$. For a two-dimensional disc it reduces to the $z$-component of vorticity divided by the surface density $\Sigma$
\begin{align}
	\zeta = \frac{\omega_{z}}{\Sigma}.
	\label{eq:vort3d}
\end{align}
For an axisymmetric background state, Eq. (\ref{eq:vort3d}) reduces to 
\begin{align}
	\zeta = \frac{1}{R \Sigma} \td{}{R} (R^2 \Omega) = \frac{\kappa^2}{2 \Sigma \Omega},
\end{align}
where $\kappa$ is the epicyclic frequency in the disc \citep{Lin2010}. In a purely Keplerian disc, the initial vortensity is given by
\begin{align} 
    \label{eq:zeta_K}
	\zeta_{0,K}(R,p) = \frac{\Omega_\mathrm{p}}{2 \Sigma_\mathrm{p}} \times \left(\frac{R}{R_\mathrm{p}}\right)^{p-3/2},
\end{align}
which is independent of $R$ for $p =3/2$. Corrections due to sub-Keplerian rotation are small for thin discs.

In barotropic two-dimensional flows that are free of shocks, the vortensity is constant along streamlines and obeys 
\begin{align}
	\frac{D \zeta}{Dt} = \pd{\zeta}{t} + \vect{u} \cdot \nabla \zeta = 0.
	\label{eq:lagzeta}
\end{align}
However, when the shocks are present $\zeta$ is no longer conserved and experiences a discontinuous jump at the shock front.

Fig. \ref{fig:dvort_map}c,d shows the vortensity deviation $\delta\zeta=\zeta-\zeta_{0}$ from its initial value\footnote{Note that $\zeta_0(R)$ is slightly different from $\zeta_{0,K}$ due to the modification of the $\Omega$ profile by the radial pressure gradient.} $\zeta_{0}$ for a simulation with fiducial disc parameters and $M_\mathrm{p} = 0.25 M_\mathrm{th}$ after 740 planetary orbits. Panel (d) shows a two-dimensional map of $\delta\zeta$ and panel (c) a radial profile of the azimuthal average. One can see that close to the planet, the initial vortensity is conserved, i.e. $\delta\zeta=0$. However, beyond the black dashed lines marking the radial separation from the planetary orbit at which the wave shocks, $|R-R_\mathrm{p}|=l_\mathrm{sh}$, the vortensity perturbation $\delta\zeta$ becomes nonzero. As fluid parcels pass through the curved shock front (seen in Fig. \ref{fig:dvort_map}a) at $|R-R_\mathrm{p}|>l_\mathrm{sh}$, they experience a small jump in vortensity; after one synodic period with respect to the planet they return to a similar position at the shock, experiencing another $\zeta$ jump of the same amplitude\footnote{We comment on long-term evolution and corrections to this simplifying assumption in Section \ref{subsec:longterm}.}, and so on. Over time, this steady accumulation of small $\Delta\zeta$ increments leads to the emergence of an almost axisymmetric vortensity distribution near the planetary orbit, exhibiting maxima (minima) close to (farther from) the planet. Emergence of the second, lower amplitude vortensity peak in the inner disc around $R \simeq 0.6 R_\mathrm{p}$ (panel (c)) is due to the shocking of a secondary spiral arms that forms prior to that (panel (a)).


\subsection{Vortensity generation by the global shock}
\label{sect:vort-model0}


We now turn to understanding the production of vortensity at the front of a planet-driven shock. For a globally isothermal EoS, the jump of $\zeta$ that a fluid parcel experiences upon crossing a shock of strength $\Delta\Sigma/\Sigma$ is given by
\citep{Kevlahan1997,Lin2010,Dong2011II}
\begin{align}
	\Delta \zeta &= \frac{\cs}{2 \Sigma \M^{5}}\left(\frac{\Delta \Sigma}{\Sigma}\right)^{2} \td{}{S} \left(\frac{\Delta \Sigma}{\Sigma}\right)\\
	&= \frac{\cs}{2 \Sigma } \frac{\left(\Delta \Sigma / \Sigma\right)^{2}}{\left(1 + \Delta\Sigma/\Sigma \right)^{5/2}} \td{}{S} \left(\frac{\Delta \Sigma}{\Sigma}\right),
	\label{eq:deltazeta}
\end{align}
where $S$ is distance measured along the shock, increasing away from the planet and we used $\M^2 = 1 + \Delta \Sigma/\Sigma$ as appropriate for the isothermal EoS, where $\M$ is the Mach number of the flow normal to the shock front. The sign of $\Delta \zeta$ is solely determined by the derivative of shock strength along the shock in this case, since all other factors are positive.

Extending the local calculation of \citet{Dong2011II}, we use Eq. (\ref{eq:deltazeta}) to derive the following analytical expression for the vortensity jump at the shock that is valid globally (see Appendix \ref{sec:thvort} for details):
\begin{align}
	\nonumber	\Delta \zeta(R) = &\frac{\cs}{2^{7/4} \Sigma_0 h_\mathrm{p}^{3/2}}  \left( \frac{M_\mathrm{p}}{M_\mathrm{th}} \right)^3 B^2(R) \left[\Delta \chi (\tau)\right]^2 \\ \nonumber
	&\times \left[ 1 + \frac{M_\mathrm{p}}{M_\mathrm{th}} \frac{B(R)}{2^{1/4} h_\mathrm{p}^{1/2}} \Delta \chi (\tau) \right]^{-5/2} \\ 
		&\times C(R) \td{}{R} \left[ B(R) \Delta \chi(\tau) \right],
		\label{eq:deltazeta2}
\end{align}
where $\tau = \tau \left(R,M_\mathrm{p}/M_\mathrm{th}\right)$. The scaling functions $B(R)$ and $C(R)$ are due to conversion between $(\Delta \Sigma/\Sigma_0)(R)$ and $\chi(\tau)$ and the geometry of the shock, respectively. We note that the term in square brackets is $\M^{-5}$ and is typically of order unity for the shock strengths considered here. 

While Eq. (\ref{eq:deltazeta2}) shows similarity to the local expression for $\Delta \zeta$ in \citet{Dong2011II}, we note that due to the nonlinear shear and the more complex shock geometry appearing in the global cylindrical disc, this formula for $\Delta \zeta$ does not reduce to an expression in terms of $\tau$ alone. However, we can still evaluate this equation numerically for different disc parameters and planet masses and compare the results directly to fully nonlinear simulations to verify the analytical prediction (\ref{eq:deltazeta2}).


\subsection{Semi-analytical model for vortensity production by the planet-driven shock}
\label{sect:vort-model}


Results of \S\ref{sec:shock} provide us with theoretical expectations for the global behavior of the shock strength $\Delta\Sigma/\Sigma_0$ and the shock position as functions of disc parameters and planet mass. Combining them with the calculations in \S\ref{sect:vort-model0} we can formulate a semi-analytical prescription for vortensity generation by the planet-driven shock, based on the framework of \citetalias{Rafikov2002} (see \S\ref{sect:wave-nonlin}). 

Given a set of disc and planet parameters --- $p,h_\mathrm{p}$, $M_\mathrm{p}/M_\mathrm{th}$ --- this prescription consists of the following sequence of steps:
\begin{enumerate}
    \item Use the definition (\ref{eq:tau}) to compute $\tau(R)$. 
    \item Use the fit Eq. (\ref{eq:fit}) with parameters in Table \ref{tab:fitpars} to predict shock strength $\Delta \chi$ as a function of $\tau(R)$. 
    \item Utilize an expression for the shock position $\phi_\mathrm{sh}$, which sets $C(R)$, see Eq. (\ref{eq:CR}). We will examine two approximations for $\phi_\mathrm{sh}$ in this work:
    \begin{enumerate}
    		\item A simple linear prediction for shock position, Eq. (\ref{eq:phi_lin}), giving $\phi_\mathrm{sh}(R)=\phi_\mathrm{lin}(R)$.
    		\item A more sophisticated prescription $\phi_\mathrm{sh}(R)=\phi_\mathrm{sh}^\mathrm{nonlin}(R,\tau(R))$, Eq. (\ref{eq:phinonlin}), accounting for nonlinear effects. 
    \end{enumerate}
    \item Plug the resultant expressions for $\phi_\mathrm{sh}(R)$ and $\Delta \chi(R)$ into Eq. (\ref{eq:deltazeta2}) to finally obtain the radial profile of the vortensity jump across the shock $\Delta \zeta (R)$.
\end{enumerate}

This recipe for calculating $\Delta \zeta (R)$ is tested and validated in the rest of the paper.


\subsection{Simulation results for vortensity generation in a fiducial disc model}
\label{sect:vort-sim}


\begin{figure}
\centering
\includegraphics[width=0.49\textwidth]{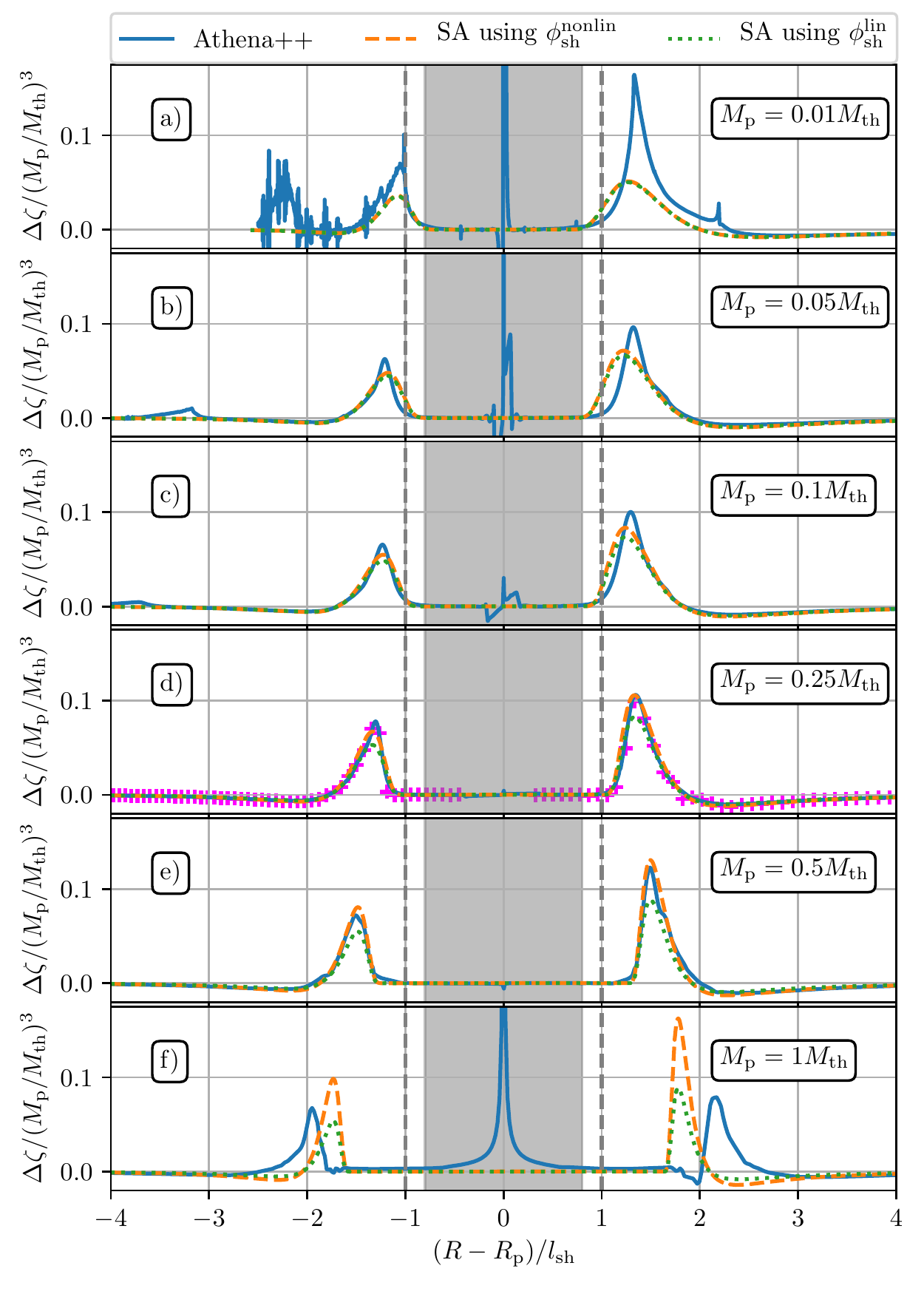}
\vspace*{-0.7cm}
\caption{Radial profiles of the vortensity jump across the shock $\Delta \zeta(R)$. We compare simulation results (blue) with semi-analytical prediction described in \S\ref{sect:vort-model}, that use different inputs for the shock location $\phi_\mathrm{sh}$: $\phi_\mathrm{sh}^\mathrm{lin}$ (green dotted) and $\phi_\mathrm{sh}^\mathrm{nonlin}$ (orange dashed). The dark grey area marks regions where vortensity
\changed{
changes, if present, are not due to the spiral shocks;
}
vertical dashed lines mark $\vert R-R_\mathrm{p}\vert = l_\mathrm{sh}$. Magenta crosses in panel (d) show $\Delta \zeta$ computed using Eq. (\ref{eq:deltazeta2}) with inputs for shock position and strength obtained directly from simulations, showing excellent agreement with observed vortensity generation. For more details see \S\ref{sect:vort-sim}.}
\label{fig:dvort-t-sim-th}
\end{figure}

\begin{figure}
\centering
\includegraphics[width=0.49\textwidth]{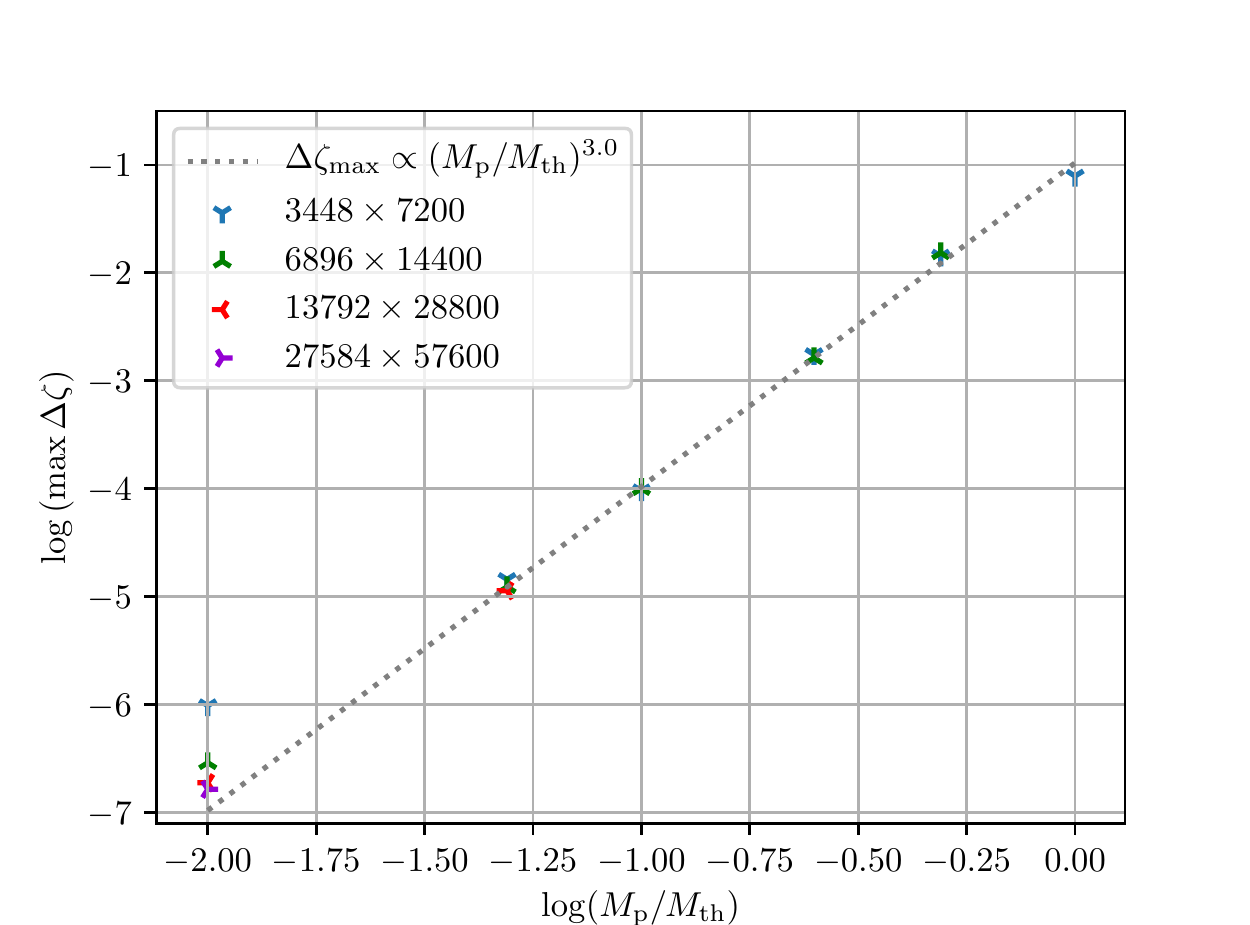}
\vspace*{-0.45cm}
\caption{Scaling of the peak value of the vortensity jump  $\Delta \zeta$ in the outer disc as a function of planet mass for a fiducial disc model. We show results of simulations with several resolutions, which demonstrate that low planet masses require high resolution to show convergence. Dotted line shows the $\Delta \zeta\propto M_\mathrm{p}^3$ scaling (see Eq. (\ref{eq:deltazeta2})) familiar from the local study of \citet{Dong2011II}.}
\label{fig:peak-dvort-scale}
\end{figure}

We now turn to vortensity generation observed in our simulations and compare it to the semi-analytical model described above.

The rate of change of vortensity per unit time is directly related to the vortensity jump $\Delta \zeta$ experienced in a single shock crossing via the synodic period:
\begin{align}
	\pd{\zeta(R)}{t} = \Delta \zeta (R) \times \frac{\vert \Omega(R) - \Omega_\mathrm{p} \vert}{2 \pi},
	\label{eq:dzetadt}
\end{align}
where $\Omega_\mathrm{p}$ appears since the shock is stationary in a frame that is corotating with the planet. We use this relation to obtain $\Delta \zeta$ from the time derivative of the azimuthally averaged vortensity, which is measured in our simulations. This time derivative is evaluated using snapshots separated by $\Delta t = 10$ planetary orbits (we found no dependence on the exact value of $\Delta t$, changing it within a few tens of orbits). For comparison with theory we perform this calculation early on, after 20-30 planetary orbits, to ensure that $\Sigma$ is unperturbed by the wave damping. We comment on the effects of long-term perturbations to the disc in \S\ref{subsec:longterm} (see also Fig. \ref{fig:dvortdt_long}). In discs with a radial vortensity gradient, radial advection of vortensity can also add to variation of $\zeta$ at fixed $R$, see Eq. (\ref{eq:lagzeta}). However, we find that in our runs the advective contribution is negligible and shock-induced vortensity generation is by far the dominant source of $\partial\zeta/\partial t$.

In Fig. \ref{fig:dvort-t-sim-th} we display the results for $\Delta\zeta(R)$ obtained from our fiducial disc model simulations (blue curves), with $M_\mathrm{p}/M_\mathrm{th}$ ranging from 0.01 to 1. Radial profiles of $\Delta\zeta$ are derived from our nonlinear simulations after 20 planet orbits using Eq. (\ref{eq:dzetadt}) in which we approximate $\partial_t \zeta(R)=\partial_t \langle\zeta(R)\rangle_\phi$. The vertical axis is rescaled by the third power of $M_\mathrm{p}/M_\mathrm{th}$, to absorb the most important scaling with $M_\mathrm{p}$ in Eq. (\ref{eq:deltazeta2}). Radial distance from the planet is rescaled by the shock distance $l_\mathrm{sh}$, which is a natural lengthscale of the problem in the local limit  \citepalias{Goodman2001}. These renormalizations lead to $\Delta\zeta(R)$ profiles that have very similar radial shape. However, slight variations are still visible as the planet mass varies, they are discussed below. 

Before the shock forms, vortensity is mostly conserved, except for numerical artifacts that are amplified for the lowest planet masses. Also, in the coorbital region of the planet we observe the well-known mixing of material on horseshoe orbits \citep{Paardekooper2009}, homogenizing the vortensity distribution.

As the density wave starts turning into a shock beyond $\vert R - R_\mathrm{p} \vert =l_\mathrm{sh}$ (marked by vertical dashed lines), a narrow positive peak in $\Delta \zeta$ forms between $1 \lesssim \vert R - R_\mathrm{p} \vert / l_\mathrm{sh} \lesssim 2$, followed by a wider but shallower negative trough at larger distances from the planet. The former is due to the rapid formation of the shock and the latter due to its subsequent decay. The overall shape of the $\Delta\zeta(R)$ curves is roughly the same in the inner and outer discs, although the outer $\Delta\zeta$ peak is higher than the inner one, illustrating the importance of global effects (in the local calculations of \citet{Dong2011II} the peaks are symmetric). Also, at small $M_\mathrm{p}$ one can see a low amplitude secondary peak in the inner disc that recedes from the planet (in $\vert R - R_\mathrm{p} \vert/l_\mathrm{sh}$ coordinate) as $M_\mathrm{p}$ increases. This peak appears due to the shocking of the secondary arm. 

Before detailed testing of the prescription outlined in \S\ref{sect:vort-model}, we checked the validity of the Eq. (\ref{eq:deltazeta2}) by using it to compute $\Delta \zeta(R)$ with the main inputs\footnote{We use a polynomial spline fit over a few cells in order to approximate the numerical parameters of the shock, which is needed to reduce the noise in computation of $\partial_S (\Delta\Sigma/\Sigma)$.} obtained directly from our simulations --- shock strength $\Delta\chi(R)$ from angular momentum flux decay, Eq. (\ref{eq:wavedamp}) and shock shape $\phi_\mathrm{sh}(R)$ --- instead of using semi-analytical fits (\ref{eq:fit}), (\ref{eq:phinonlin}), (\ref{eq:phi-pars}). This calculation is illustrated for  $M_\mathrm{p}=0.25 M_\mathrm{p}$ by magenta crosses in Fig. \ref{fig:dvort-t-sim-th}d, providing an excellent match with simulation results and confirming that the shock is the only relevant source of vortensity. This agreement also shows that Eq. (\ref{eq:deltazeta2}) is valid and the code performs as expected.

Green dotted and orange dashed curves in Fig. \ref{fig:dvort-t-sim-th} illustrate the semi-analytical prescriptions for $\Delta \zeta (R)$ (described in \S\ref{sect:vort-model}), computed for $\phi_\mathrm{sh}=\phi_\mathrm{lin}$ and $\phi_\mathrm{sh}=\phi_\mathrm{sh}^\mathrm{nonlin}$, respectively. By design, these prescriptions agree best with simulation results for $0.1 \leq M_\mathrm{p}/M_\mathrm{th} \leq 0.5$, since the parameters (see Table \ref{tab:fitpars}) of the fit (\ref{eq:fit}) were determined for $M_\mathrm{p}=0.25 M_\mathrm{th}$. The discrepancies at low $M_\mathrm{p}$ arise most likely because of the increased importance of linear (validity of the fit for $\Delta \chi$) and numerical effects for small-amplitude density waves (see the discussion of Fig. \ref{fig:peak-dvort-scale} below).

In the highest mass case ($M_\mathrm{p}=M_\mathrm{th}$), the agreement with the simulation results is poor, since nonlinear effects are strong and the shock forms while the wave is still accumulating angular momentum, in contrast to the assumptions of the \citetalias{Goodman2001} and \citetalias{Rafikov2002}, see \S\ref{sect:wave-nonlin}. This spatial overlap of the excitation and propagation regions is also responsible for the shift of vortensity peaks away from $\vert R-R_\mathrm{p}\vert=l_\mathrm{sh}$ at high $M_\mathrm{p}$. This is because in the global case the rescaling of radial separation by $l_\mathrm{sh}$ is not identical to expressing the distance from the planet through the coordinate $\tau$.

Semi-analytical curves computed for the two different approximations for $\phi_\mathrm{sh}$ (see \S\ref{sect:vort-model}) differ from each other only for $M_\mathrm{p}\gtrsim 0.5M_\mathrm{th}$. This is expected, since at low masses nonlinear effects are weak and 
$\phi_\mathrm{sh}^\mathrm{nonlin}\approx \phi_\mathrm{lin}$. We find that, for practical purposes, using the simple approximation $\phi_\mathrm{sh}=\phi_\mathrm{lin}$ for semi-analytical calculation of $\Delta\zeta(R)$ is at least not inferior to the more sophisticated assumption $\phi_\mathrm{sh}=\phi_\mathrm{sh}^\mathrm{nonlin}$. 

Fig. \ref{fig:dvort-t-sim-th} clearly illustrates that over a wide range of planet masses $(0.05 - 0.5 M_\mathrm{th})$, vortensity generation exhibits the self-similar behavior expected in the framework of \citetalias{Rafikov2002}, as long as the distance from the planet and $\Delta\chi$ amplitude are scaled according to theory. The vortensity amplitude scaling is additionally illustrated in Fig. \ref{fig:peak-dvort-scale}, where we show the peak value of $\Delta \zeta$ in the outer disc as a function of planet mass in simulations using the fiducial disc model. One can see a clear power law behavior scaling that is well fitted with max$\Delta \zeta \propto (M_\mathrm{p}/M_\mathrm{th})^3$  (dotted line in the figure). This scaling was found in \citet{Dong2011II} in the local (shearing-sheet) approximation, but clearly is valid in the global case as well. This is not surprising since $\Delta\zeta\propto M_\mathrm{p}^3$ is the dominant dependence on $M_\mathrm{p}$ in the global Eq. (\ref{eq:deltazeta2}), with additional $M_\mathrm{p}$-dependent terms playing a minor role.  

Fig. \ref{fig:peak-dvort-scale} also illustrates the numerical convergence of our results over a wider range of resolutions, indicated by different symbols. One can see that convergence is very good, except at the lowest $M_\mathrm{p}=0.01M_\mathrm{th}$, which requires the highest resolution (464 cells per scale-height at the planets orbital radius) to appear converged.


\subsection{Effect of gap opening}
\label{subsec:longterm}


\begin{figure}
\centering
\includegraphics[width=0.49\textwidth]{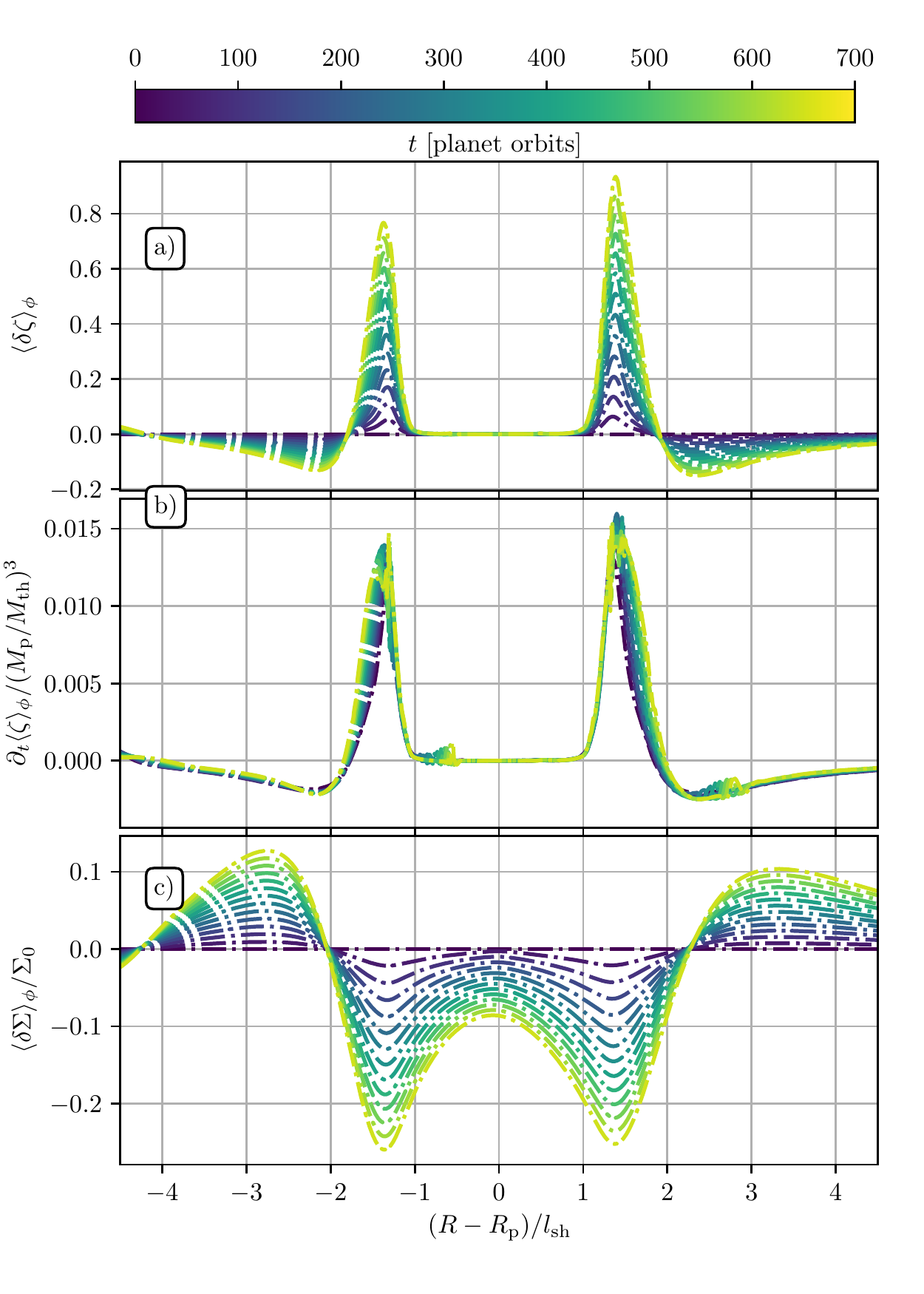}
\vspace*{-0.8cm}
\caption{Time series of radial profiles of (a) vortensity perturbation, (b) its time derivative (proportional to $\Delta\zeta$), and (c) surface density perturbation, illustrating the effect of a forming gap. The results are shown for $M_\mathrm{p}/M_\mathrm{th} = 0.25$ and the fiducial disc parameters, however using the lowest resolution in order to show evolution over a longer time-scale.}
\label{fig:dvortdt_long}
\end{figure}

Up to this point we always assumed the disc surface density near the planet to be unperturbed and vary only on scales of order $R_\mathrm{p}$. However, as the angular momentum lost by the damping density wave is absorbed by the disc fluid, a radial redistribution of gas around the planet's orbit takes place, eventually resulting in gap opening. This will modify the planet-driven wave in several ways. 

First, suppression of the surface density near the planet reduces the strength of its gravitational coupling to the disc, resulting in weaker planetary torque and lowering the density wave amplitude at excitation \citep{Petrovich2012}. Second, as the wave propagates across an inhomogeneous background at the gap edge, its nonlinear steepening and decay get modified as described in \citetalias{Rafikov2002} and \citet{Rafikov2002II}. These factors combine to modify vortensity generation by the planet-driven shock.

We illustrate this in Fig. \ref{fig:dvortdt_long} by showing a time-series (over 700 orbits) of azimuthal averages of the vortensity perturbation (panel a), its time-derivative (panel b), and the surface density profile (panel c). Results are shown for the fiducial disc model and $M_\mathrm{p} = 0.25 M_\mathrm{th}$. 

One can see that a characteristic double gap \citep{Rafikov2002II} slowly grows in depth around the planetary orbit. This process is accompanied by   continuous enhancement of the vortensity perturbation, with the positive (negative) $\delta\zeta(R)$ associated with reduced (increased) surface density. This steady increase of  $\delta\zeta$ eventually produces large radial gradients of vortensity near the planet that can lead to non-negligible radial {\it advective} transport of vortensity (which we normally find to be unimportant) during gap opening.

At early times, $\partial_t \langle \delta \zeta \rangle_\phi$ exhibits a very stable, time-independent profile (that we use in Fig. \ref{fig:dvort-t-sim-th}). But once gap depth reaches $\langle \delta \Sigma \rangle_\phi / \Sigma_0 \lesssim 0.15$, we find $\partial_t \langle \delta \zeta \rangle_\phi$ to deviate from the initial smooth profile. At this point, since $\delta\Sigma$ is still relatively small, we observe only localized changes of $\partial_t \langle \delta \zeta \rangle_\phi$ profile: its peak broadens and goes down in amplitude, also showing a small kink around the maximum value (most likely due to vortex formation at late times). On longer timescales, when the gap gets deeper, we expect vortensity production at the shock to be modified more severely. 


\section{Results: Variation of the disc parameters}
\label{sec:resdisc}


Next we examine how our results and the comparison with semi-analytical calculations change as we vary the underlying disc model. We will predominantly focus on the AMF behavior and vortensity generation as our metrics for comparison since they provide very sensitive diagnostics of the nonlinear wave evolution.


\subsection{Variation of the surface density slope}
\label{sect:surf_var}


\begin{figure}
\centering
\includegraphics[width=0.49\textwidth]{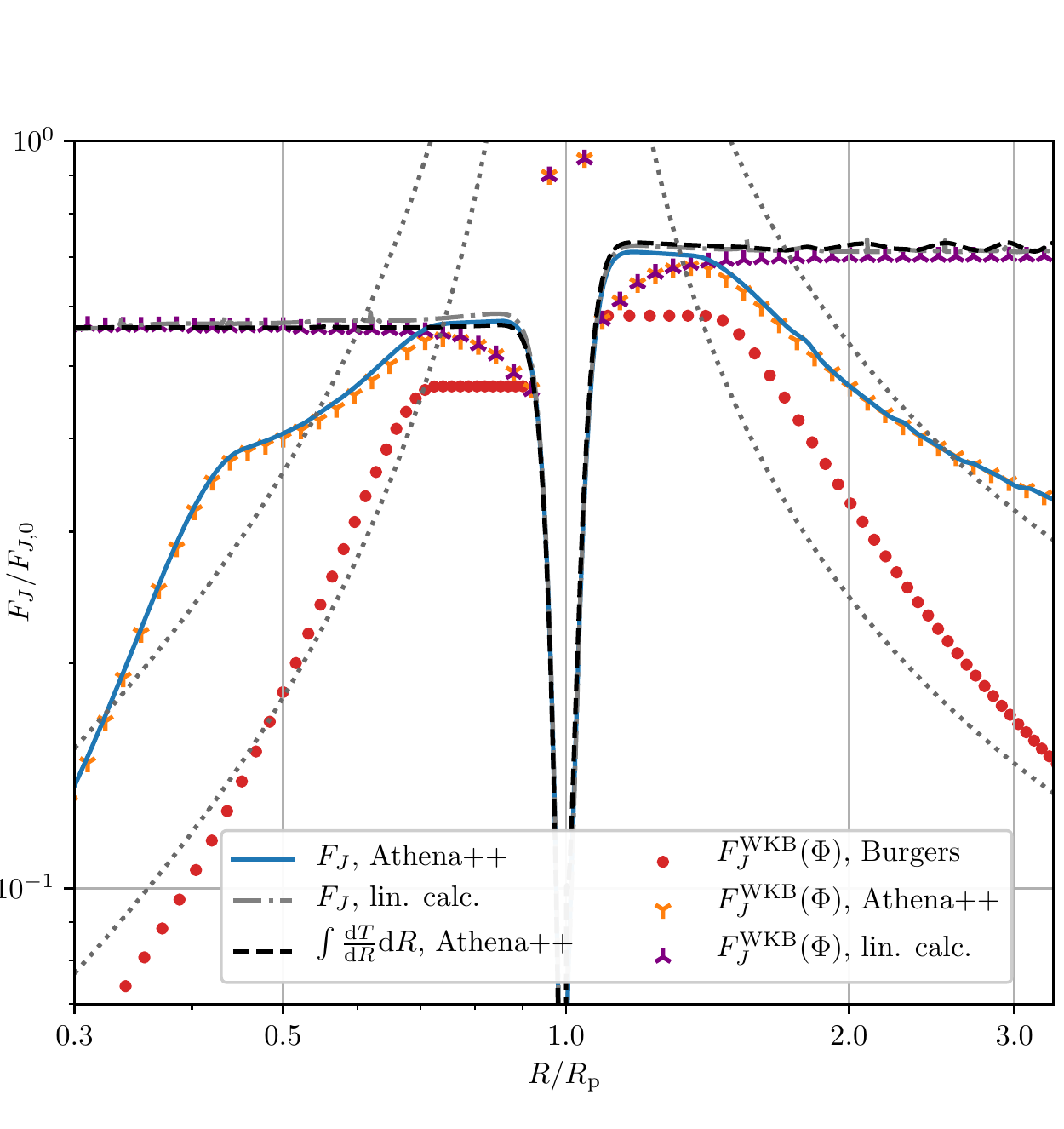}
\vspace*{-0.7cm}
\caption{Same as Fig. \ref{fig:AMF001}, i.e. $M_\mathrm{p}=0.01M_\mathrm{th}$, but for the uniform disc case ($p=0$). As expected directly from the dependence of the variable $\tau$ on $p$ (Eqs. (\ref{eq:tau_g})-(\ref{eq:g_g}) and Fig. \ref{fig:tau-r}), wave damping is increased (decreased) in the inner (outer) disc as compared to the fiducial disc. See text for details.}
\label{fig:AMF001q0}
\end{figure}

\begin{figure}
\centering
\includegraphics[width=0.49\textwidth]{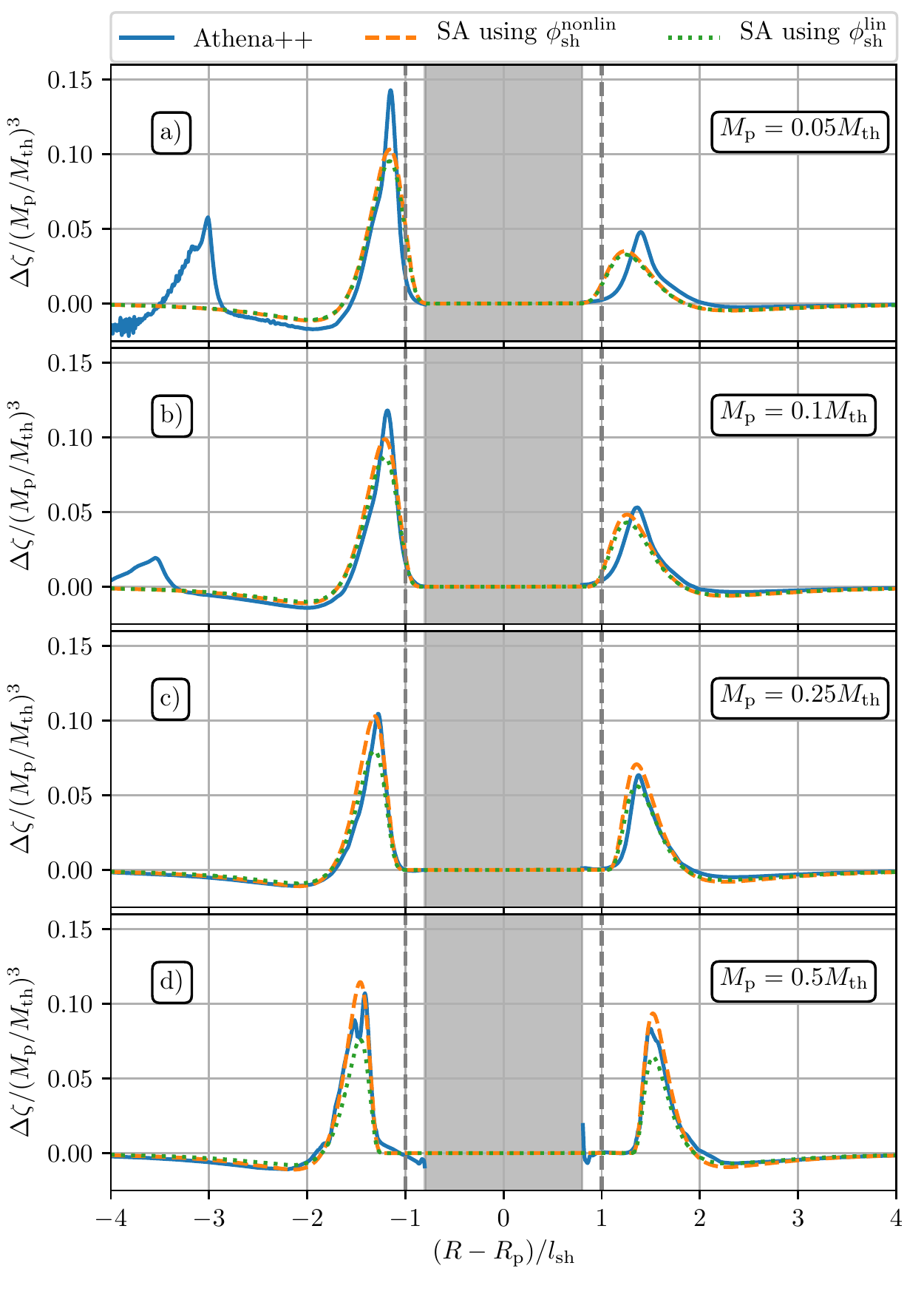}
\vspace*{-.65cm}
\caption{Same as Fig. \ref{fig:dvort-t-sim-th} but for constant surface density disc ($p=0$). Note that, compared to the $p=3/2$ disc (see Fig. \ref{fig:AMF001}), the peak of $\Delta\zeta$ is now higher in the inner disc.
\changed{
We omit simulation data in the corotation region (marked grey), where vortensity changes are due to advective mixing of material.
}
}
\label{fig:q0-dvort-t-sim-th}
\end{figure}

First, we turn to a disc model that has the fiducial value of $h_\mathrm{p} = 0.05$ but a {\it constant} background surface density, i.e. $p=0$, see Eq. (\ref{eq:model-PLs}). Different from the fiducial disc, this model has a global non-zero vortensity gradient, since its $\zeta_{0,K}(p=0) \propto \left( R/R_\mathrm{p} \right)^{-3/2}$, see Eq. (\ref{eq:zeta_K}). This may somewhat increase the role of the advective transport of vortensity. Given the generality of the framework outlined in \S\ref{sect:wave-nonlin}, we expect the effects of a different slope of $\Sigma_0(R)$ on wave evolution to be fully absorbed in the definitions of the coordinates $\chi$ and $\tau$.

We begin by presenting in Fig. \ref{fig:AMF001q0} the behaviour of AMF for a planet of mass $M_\mathrm{p} = 0.01M_\mathrm{th}$. Comparing it to Fig. \ref{fig:AMF001}, we see very similar behaviour close to the planet: after wave excitation, its AMF reaches values comparable to the fiducial case and begins to decay at a similar distance. Thus, the shocking length $l_\mathrm{sh}$ is still captured by the Eq. (\ref{eq:lsh}). Moreover, the $R-\phi$ shape of the wake is still accurately fit by the Eqs. (\ref{eq:phinonlin})-(\ref{eq:phi-pars}).

However, on the global scale, we see that wave damping is accelerated (slowed down) in the inner (outer) disc. As a consequence less (more) AMF reaches the damping boundary. This is true for the full simulation as well as for the solutions of the Burgers equation and can
be directly understood from the effect of $p$ on $\tau$: according to Fig. \ref{fig:tau-r}, $\tau$ reaches higher (lower) values in the inner (outer) disc, as compared to the fiducial disc. This behaviour is observed across all planet masses. Qualitative differences between evolution described by the full set of hydro equations (Athena++) and by Burgers equation (see \S\ref{sec:res}) still remain.

Next we examine the vortensity generation. Our semi-analytical framework should provide a good match for the $\zeta$ production at the shock when $p=0$, since all variables in our framework naturally account for arbitrary $p$. In Fig. \ref{fig:q0-dvort-t-sim-th}, we compare simulation results against the semi-analytical model (\S\ref{sect:vort-model}) for $p=0$.
\changed{
The filled grey area marks the corotation region, where the horseshoe flow mixes regions of
different initial vortensity advectively. We do not show simulation data in this region as $\zeta$ changes there are not due to shocks.
}
First we note that, as opposed to the fiducial disc, the inner peak now {\it dominates} over the outer one in the linear regime ($M_\mathrm{p}/M_\mathrm{th}=0.01$).
Very importantly, this behaviour is correctly captured by our semi-analytical model (orange dashed curve). For the three higher planet masses ($0.1 \leq M_\mathrm{p}/M_\mathrm{th} \leq 0.5$), the agreement between our semi-analytical theory and simulations is very good.


\subsection{Variation of the disc scale-height}
\label{sect:h_var}


\begin{figure}
\centering
\includegraphics[width=0.49\textwidth]{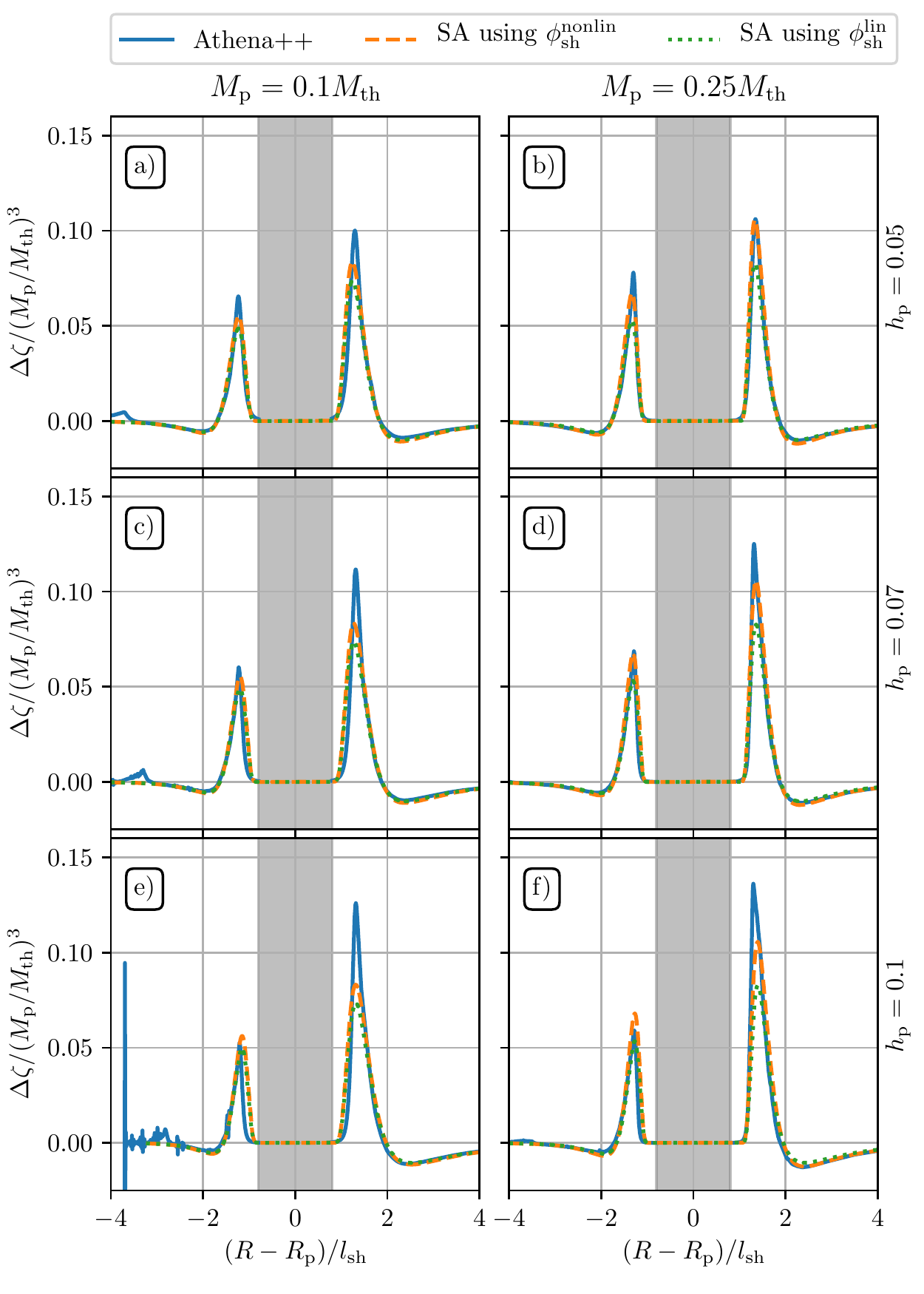}
\vspace*{-0.8cm}
\caption{Same as Fig. \ref{fig:dvort-t-sim-th} but now for three values of the disc aspect ratio $h_\mathrm{p}$ (constant across rows) and two values of $M_\mathrm{p}/M_\mathrm{th}$ (constant within columns). In the bottom left panel, the inner damping zones and domain boundary lies within $4l_\mathrm{sh}$, which results in somewhat noisy data.}
\label{fig:THvort_hvar}
\end{figure}

Next we explore the effect of variation of the disc aspect ratio by studying simulations with $h_\mathrm{p} = 0.07$ and $0.1$. These are performed at the same resolution as the fiducial models that we deem converged, effectively increasing the number of grid cells per scale-height.
\changed{
Thus, simulations at larger $\hp$ are expected to be converged too, which we have confirmed.
}

Variation of $h_\mathrm{p}$ is expected to have a direct effect on the amplitude of the wave perturbation at excitation. For a fixed planet mass, higher $h_\mathrm{p}$ results in smaller amplitude of the perturbation (since $M_\mathrm{th}$ is higher and $M_\mathrm{p}/M_\mathrm{th}$ is lower) and the density wave carries less angular momentum, see Eq. (\ref{eq:FJ0}). But if we keep $M_\mathrm{p}/M_\mathrm{th}$ fixed , then the wave amplitude, expressed through $\chi$, remains the same. From Eq. (\ref{eq:tau}), we also find $\tau \propto h_\mathrm{p}^{-5/2}$, meaning that nonlinear evolution, expressed via $\tau$-coordinate, is slowed down in hotter discs. In terms of radial distance, this dependence is approximately absorbed in the dependence of $l_\mathrm{sh}$ on $h_\mathrm{p}$, as we demonstrate below. And regarding the $R-\phi$ shape of the wake (\S\ref{sec:phishock}), we find that the linear dependence on $h_\mathrm{p}$ of the nonlinear correction term $\phi_\mathrm{sh}^\mathrm{nonlin}-\phi_\mathrm{lin}\propto h_\mathrm{p}$ in Eq. (\ref{eq:phinonlin}) provides an excellent match to the simulation results.

In Fig. \ref{fig:THvort_hvar} we show $\Delta \zeta$ as a function of $h_\mathrm{p}$ (rows) for $M_\mathrm{p}/M_\mathrm{th} = 0.1$ (left column) and $M_\mathrm{p}/M_\mathrm{th} = 0.25$ (right column). For both masses, the general shape of $\Delta \zeta$ is maintained and $\Delta\zeta$ curves appear very similar when $R-R_\mathrm{p}$ is rescaled by $l_\mathrm{sh}$. Thus, our semi-analytical model (\S\ref{sect:vort-model}) provides a good match to simulation results. 

Note that as $h_\mathrm{p}$ increases, peak values of $\Delta\zeta$ increase (decrease) in the outer (inner) disc. We traced this behavior to the effect of $h_\mathrm{p}$ on the initial (linear) wake profile: for higher $h_\mathrm{p}$ we find the magnitude of minimum and maximum values of $\chi$ to increase (decrease) in the outer (inner) disc. But the main effect of changing $h_\mathrm{p}$ is the variation of the spatial scale of the problem. Since the rescaled (horizontally, by $l_\mathrm{sh} \propto h_\mathrm{p}$) profiles of $\Delta \zeta$ in Fig. \ref{fig:THvort_hvar} remain largely unchanged, this implies that in physical space (in $R-R_\mathrm{p}$) the vortensity profile is broader for higher $h_\mathrm{p}$, see Fig. \ref{fig:dchi-theory} for illustration.


\section{Discussion}
\label{sec:disc}


The main goals of this work were to numerically verify the semi-analytical theory of planet-driven density wave propagation advanced in \citetalias{Rafikov2002}, and to explore the impact of the emergence of secondary (and higher order) spiral arm on wave evolution in the inner disc. Using high-resolution Athena++ simulations we were able to confirm the main predictions of the semi-analytical framework of \citetalias{Rafikov2002} summarized in \S\ref{sect:wave-nonlin}.

In particular, we find that the change of coordinates from $R$, $\phi$, $\Sigma-\Sigma_0$ to $\tau$, $\eta$, $\chi$ works very well when describing the planet-driven density wave evolution, as long as certain problem-specific inputs are properly calibrated using simulations. Examples of this include the scaling of $\phi_\mathrm{sh}^\mathrm{nonlin}-\phi_\mathrm{lin}$ with $h_\mathrm{p}$ and $\tau$ (see Eq. (\ref{eq:phinonlin}) in \S\ref{sec:phishock}), self-similarity of the shock strength profile (see Eq. (\ref{eq:fit}) in \S\ref{sec:deltachi}), semi-analytical calculation of the vortensity production in \S\ref{sect:vort-model}, and so on. In agreement with the findings of \citet{Dong2011II} and \citet{Duffell2012}, we find very good agreement between the theoretically predicted shocking distance $l_\mathrm{sh}$ (see Eq. \ref{eq:lsh}) and our global simulation results, even though $l_\mathrm{sh}$ was derived in \citetalias{Goodman2001} in the local (shearing sheet) approximation. The evolution of the density wave profile towards the N-wave shape is also reproduced in our simulations, see Figs. \ref{fig:lin-nonlin}, \ref{fig:025nonlin-burger}.

Using vortensity generation as a metric, the best agreement between the (calibrated) theory and simulations is found for intermediate planet masses $M_\mathrm{p} \simeq 0.05-0.5M_\mathrm{th}$. At higher masses approaching $M_\mathrm{th}$ the agreement worsens (see Fig. \ref{fig:dvort-t-sim-th}f), since the planet-driven wave is nonlinear already at excitation and the key theoretical assumption of spatial separation between linear excitation and nonlinear propagation breaks down. At low $M_\mathrm{p} < 0.05M_\mathrm{th}$, numerical effects as well as the proximity of the wave damping zones (because of the increased $l_\mathrm{sh}$, see Eq. (\ref{eq:lsh})), also worsen the agreement with theory, see Fig. \ref{fig:dvort-t-sim-th}a.

The emergence of the secondary spiral arm in the inner disc certainly affects the performance of the \citetalias{Rafikov2002} theory, which did not take this subtle linear effect into account. In the inner disc, and especially at low $M_\mathrm{p}$, that theory (as well as its local analogue, see \citetalias{Goodman2001}) misses the appearance of a second bump in the profile of $\Delta\chi$ at the shock as a function of $\tau$ (see Fig. \ref{fig:dchi-t}), slow decay of the AMF $F_J$ in the inner disc (see Fig. \ref{fig:AMF001}), and secondary peaks of vortensity production (see Fig. \ref{fig:dvort-t-sim-th}). For these reasons, in the inner disc the semi-analytical framework of \citetalias{Rafikov2002} is valid only for $R\gtrsim (0.5-0.6) R_\mathrm{p}$, before the secondary arm fully forms.


\subsection{Validity of Burgers equation for describing evolution of  planet-driven density waves}
\label{ssec:globwave}


While, as described above, many predictions of the semi-analytical theory of \citetalias{Rafikov2002} work very well, once properly calibrated by simulations, some other aspects of theory, starting from first principles, show only qualitative agreement in many cases. In particular, when we attempt to reproduce wave evolution by directly solving the Burgers equation (rather than using simulations to calibrate semi-analytical prescriptions), we find quantitative differences with the simulations. They can be clearly seen in Fig. \ref{fig:lin-nonlin}, which reveals a faster decay and slower azimuthal spreading of the density wave profile evolved using Burgers equation (\ref{eq:Burger}), as compared to the simulation results, for a low mass planet. Unsurprisingly, this leads to a considerably faster decay of the Burgers $F_J^\mathrm{WKB}$ compared to both $F_J$ and $F_J^\mathrm{WKB}$ in simulations, see Fig. \ref{fig:AMF001}.     

\begin{figure*}
\centering
\includegraphics[width=0.99\textwidth]{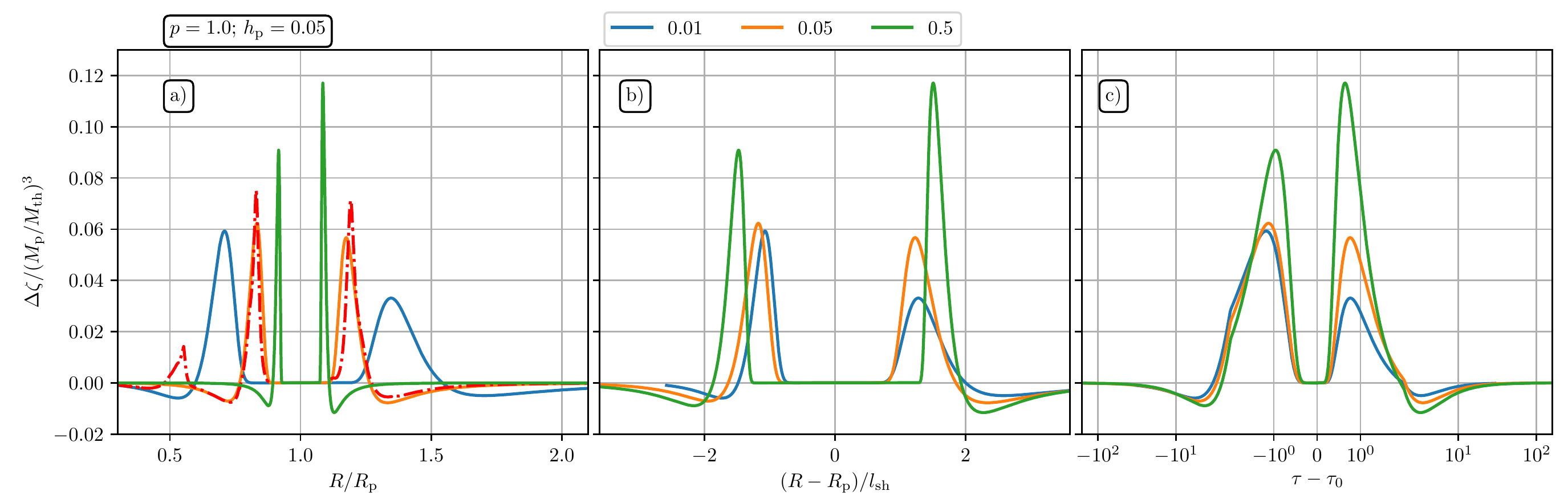}
\vspace*{-0.2cm}
\caption{
Semi-analytical predictions for vortensity jump in inner and outer disc (solid lines) for a disc with $p=1$ and $\hp = 0.05$. We show results for
three values of $\Mp/\Mth$, as functions of the different coordinates: orbital stellocentric radius (panel a), distance from the planet normalized by $l_\mathrm{sh}$ (panel b) and the coordinate $\tau - \tau_0$ (panel c). In panel (a), we additionally show results from a simulation for $\Mp/\Mth = 0.05$ (red dot-dashed curve), which confirms that the $\Delta\zeta$ peaks in the inner and outer disc are of almost equal height for these disc and planet parameters, as predicted by our semi-analytical model (see \S\ref{sect:vort-discussion} for details).
}
\label{fig:dchi-theory}
\end{figure*}

This may seem surprising given that the Burgers equation (\ref{eq:Burger}) uses the simulated wake profile near the planet as a starting point for its subsequent evolution. This implies that Burgers equation is not capturing well some wave propagation physics that takes place in the actual simulation. We believe that, at least partly, this missing ingredient is the continued injection of the angular momentum into the density wave by the planetary potential happening in simulations far from the planet, outside of the nominal excitation region. Since the wave amplitude decays due to its dissipation at the shock, addition of even a small amount of angular momentum to the wave far from the planet can significantly slow down the decay of $F_J$ in simulations. By design, this injection is absent in the Burgers equation approach, resulting in the discrepant $F_J^\mathrm{WKB}$. An indirect support for this possibility can be seen in the less discrepant $F_J^\mathrm{WKB}$ for higher $M_\mathrm{p}=0.25M_\mathrm{th}$ planet, see Fig. \ref{fig:AMF025}: due to faster decay of the wave amplitude in the higher $M_\mathrm{p}$ simulation, less angular momentum gets added to the wave far from the planet by its torque.  

To test this hypothesis in a more direct way we performed an additional $M_\mathrm{p}=0.25M_\mathrm{th}$ simulation in which we artificially suppressed the planetary potential beyond some truncation radius $d_\mathrm{t}$, thus preventing the continuing injection of angular momentum flux into the wave far from the planet. This is one way in which a freely propagating density wave can be realized (cf. \citealt{Arza2018}), bringing wave evolution closer to the regime described by the Burgers equation. In practice, we modify the potential (\ref{eq:phi4}) using a cut-off function, such that	$\Phi_\mathrm{p}(d) \rightarrow f(d) \times \Phi_\mathrm{p}(d)$, where $d$ is the distance from the planet, 
\begin{align}
	f(d) = \begin{cases}
	1 &\quad \text{for } d \leq d_\mathrm{t}, \\
	A \exp \left( -\vert d-d_\mathrm{t} \vert/w\right) &\quad \text{for } d>d_\mathrm{t},
	\end{cases}
	\label{eq:phi-trunc}
\end{align}
$d_\mathrm{t} = 0.11 R_\mathrm{p} \simeq 2.2 H_\mathrm{p}$ is the truncation distance, $w = 0.01R_\mathrm{p}$ is the characteristic scale for a potential decay, and $A$ is a normalization constant making the resulting potential continuous.

The AMF behavior resulting in this run is shown in Fig. \ref{fig:AMF025} via green solid curve ($F_J$) and brown crosses ($F_J^\mathrm{WKB}$), see the legend. One can see that around excitation, the original simulation and the one with truncated potential predict very similar AMF behavior. But further out, the AMF for a truncated potential decays faster than for the non-truncated $\Phi_\mathrm{p}$, which can be see by comparing blue and green curves. Both $F_J$ and $F_J^\mathrm{WKB}$ for a truncated potential stay closer to the Burgers $F_J^\mathrm{WKB}$ (red points). This provides support to our hypothesis that it is the distant gravitational coupling of the density wave to the planetary potential, that causes the Burgers equation to overpredict the wave decay. On the other hand, the green curve does not fully converge to the red points, which suggests that some other factors may also be at play.  

To summarize, while the Burgers equation (\ref{eq:Burger}) provides a useful framework for understanding the planet-driven wave evolution at the qualitative level, a more accurate approach (e.g. a theory calibrated by simulations) is needed for quantitative agreement with simulations. This can be important for certain applications, see \citet{Bollati2021} and \S\ref{sect:applications}.


\subsection{Vortensity generation at the shock and its semi-analytical description}
\label{sect:vort-discussion}


Our derivations in \S\ref{sect:vort-model} and Appendix \ref{sec:thvort} provide a natural generalization of the local (i.e. shearing sheet geometry) calculation of the vortensity production in \citet{Dong2011II} to the case of a global (i.e. polar) disc geometry with radial gradients of $\Sigma$. Our calculation of the vortensity generation at the shock can also naturally account for the modification of the global shape of the spiral shock by the nonlinear effects (\S\ref{sec:phishock}), although we find the resulting corrections to be significant only for high $M_\mathrm{p}$, see \S\ref{sect:vort-sim} and Fig. \ref{fig:dvort-t-sim-th}.

This calculation provides a basis for the semi-analytical framework for predicting the vortensity production at the shock, described in \S\ref{sect:vort-model}. This framework employs a single input --- the jump of the dimensionless wave perturbation $\Delta\chi$ at the shock --- that is calibrated using simulations. This calibration step yields an accurate description of $\Delta\zeta(R)$, that compares very well against simulations, see \S\ref{sect:vort-sim}, \ref{sec:resdisc}. This, in particular, implies that the radial transport of vortensity has negligible effect on its evolution in the vicinity of the planet, i.e. the variation of $\zeta$ can be explained entirely through its production by the planet-driven shock.

The semi-analytical framework for vortensity production is one of the main results of our work and will be used in future calculations of the planet-induced vortensity evolution. Its accuracy allows us to make inferences about the vortensity evolution without running computationally expensive hydrodynamical simulations. 

For example, we used this framework to determine the value of the surface density slope $p$ at which the height of the vortensity peaks in the inner and outer disc would be the same; this question is relevant for determining the side of the disc in which vortices triggered by the planetary perturbation might first appear
\changed{
(Cimerman \& Rafikov, in prep.).
}
Our semi-analytical approach allows a very efficient exploration of this problem. Given that we previously found the inner (outer) peak of $\Delta\zeta$ to dominate for $p=0$ ($p=3/2$), one should expect the transition (equal amplitude of the vortensity peaks) to take place for some $p$ in the interval $(0,3/2)$. However, it turns out that the value of $p$ at which this transition takes place also depends on the planet mass. We demonstrate this in Fig. \ref{fig:dchi-theory} where we display several semi-analytical $\Delta\zeta(R)$ profiles computed for $p=1$, for three values of $M_\mathrm{p}$ and as functions of different coordinates: $R/R_\mathrm{p}$, $(R-R_\mathrm{p})/l_\mathrm{sh}$ and $\tau(R)$. One can see that for this value of $p$ the inner peak dominates for low $\Mp$ (blue curve), while for high $\Mp$ the outer vortensity peak is higher (green curve). Our theory predicts that $\Delta\zeta$ peaks should be almost equal in amplitude for $\Mp = 0.05 \Mth$. And indeed, in panel (a), we also show data from a simulation run for $\Mp = 0.05 \Mth$ and $p=1$ (red dashed) that agrees very well with this theoretical prediction (orange solid).

This figure conveniently illustrates several other important points about the vortensity evolution: scaling of $\Delta\zeta$ with $M_\mathrm{p}$, (almost) self-similar appearance of $\Delta\zeta$ when expressed in  $\tau(R)$ and $(R-R_\mathrm{p})/l_\mathrm{sh}$ coordinates, and the increase of radial scale of $\Delta\zeta$ variation as $M_\mathrm{p}$ is decreased, see panel (a).


\subsection{Effect of equation of state and disc thermodynamics}
\label{sect:EoS}


Our calculations are restricted to globally isothermal discs,. At the next level of sophistication, one might wonder what changes would be brought in by considering e.g. the {\it locally} isothermal EoS, in which $c_s$ is a function of $R$. We expect two modifications to be introduced by this EoS. 

First, \citet{Lin2010} showed that in a locally isothermal disc vortensity jump across a shock differs from equation (\ref{eq:deltazeta}) by an additional (baroclinic) term:
\begin{align}
	\Delta \zeta = \frac{\cs}{2 \Sigma } \frac{\left(\Delta \Sigma / \Sigma\right)^{2}}{\left(1 + \Delta\Sigma/\Sigma \right)^{5/2}} \td{}{S} \left(\frac{\Delta \Sigma}{\Sigma}\right) + \frac{2\Delta \Sigma}{\Sigma^2 (\Delta \Sigma/\Sigma + 1)^{3/2}} \td{c_{s}}{S}.
\end{align}
\citet{Lin2010} have also argued that this second (baroclinic) term should often be negligible, since $c_s$ varies on scales of order the disc size, whereas $\Delta\Sigma$ across the shock varies over shorter length scales near the planet. Thus, we expect this term to be unimportant for high mass planets, for which  $l_\mathrm{sh}\ll R_\mathrm{p}$. But for slowly decaying, lower amplitude waves driven by the low mass planets ($\Mp\ll \Mth$), this term might contribute to vortensity generation at the shock\footnote{Additional baroclinic vortensity generation may occur during evolution of the flow, for example in the corotation region, where horseshoe orbits approach the planet \citep{Paardekooper2008} or later on during the onset of the Rossby Wave Instability \citep{Lovelace1999}.} more strongly.

Second, the behavior of $\Delta\Sigma$ is different in locally and globally isothermal discs. Part of the variation comes from the explicit difference in temperature profiles near the planet, but there is also a more subtle contribution. As shown in \citet{Miranda2019II}, angular momentum of the density waves propagating in locally isothermal discs is {\it not conserved}, unlike in globally isothermal ones. Instead, it scales with the local disc temperature, i.e. $\propto c_s^2$, as a result of additional coupling with the background flow (manifesting itself even in the linear regime). This translates into a different behavior of $\Delta\Sigma(R)$ compared to the discs considered in this work.

We have explored evolution of the wave angular momentum flux in a locally isothermal disc using FARGO3D and Athena++. We found that, especially for sub-$\Mth$ planets, wave damping is accelerated or slowed down substantially,
beyond the expected effect of changing $\tau$, in agreement with the findings of \citet{Miranda2019II}. This changes the $\Delta\Sigma$ and $\Delta \chi(R)$ behaviour in a way that is not obviously generalizable within our current framework. The same is true in the even more general case of a disc with explicit heating/cooling, in which angular momentum flux of planet-driven density waves is known to not be conserved in general \citep{Miranda2020I,Miranda2020II,Zhang2020}.


\subsection{Limitations of this work}
\label{sect:limitations}


Our work necessarily employs a number of simplifications, in addition to the ones regarding disc thermodynamics (\S\ref{sect:EoS}). One of them is our assumption that the planet-driven density waves decay only as a result of irreversible dissipation at the shock, into which they evolve due to nonlinear effects. In real discs there are other, linear, mechanisms that can be responsible for wave damping. In particular, radiative damping has been demonstrated recently \citep{Miranda2020I} to play a strong role in wave dissipation in protoplanetary discs. Especially for lower mass planets, radiative losses can easily dominate wave dissipation compared to the nonlinear damping at the shock \citep{Miranda2020II}. While accounting for the radiative wave damping is possible in principle, as described in \citet{Miranda2020I,Miranda2020II}, in practice this procedure may be too cumbersome, leaving direct simulations with explicit heating/cooling as a better option. 

We have also neglected the effects of viscosity in our study. While viscous damping of density waves is typically subdominant compared to either linear radiative or nonlinear effects \citep{Miranda2020II}, viscosity can smooth out the emerging density gradients and affect the vortensity evolution. However, recent studies \citep{Rafikov2017,Flaherty2017,Flaherty2018,Flaherty2020} have shown that effective viscosities in protoplanetary discs are likely low.

We also neglected the possibility of planet migration, which should result from the asymmetry of the torques exerted by the planet on the disc. Migration can produce additional asymmetry of the surface density and torque distribution around the planet \citep{Rafikov2002II}. In this regard, our results should still be valid on time-scales over which
the planet would not migrate significantly.

Finally, our 2D study neglects the possibility of vertical motions in the disc and other related 3D effects. However, previously \citet{Zhu2015} have shown that many aspects of the density wave propagation in 2D discs, including the nonlinear effects, directly translate into fully 3D discs (see next).


\subsection{Comparison to previous work}
\label{sect:comparison}


Some aspects of the nonlinear density wave evolution have been studied numerically in the past. In particular, \citet{Duffell2012} looked at some properties of the global planet-driven spiral arms using high-resolution 2D simulations. We find good qualitative agreement with the behaviour they reported for $F_J$ derived the WKB approximation, see Eqs. (\ref{eq:FJPhi})-(\ref{eq:Phi_def}). Comparing our lowest mass case ($M_\mathrm{p}/M_\mathrm{th} = 0.01$) to theirs ($M_\mathrm{p}/M_\mathrm{th} = 0.0209$), we find very good agreement in the inner disc, including the effect of the secondary arm formation on $F_J$, which is now well understood. In the outer disc, our results show a slower than $F_J \propto \tau^{-1/2}$ decay with $\tau$, whereas \citet{Duffell2012} find a behaviour close to this scaling. We speculate that his might be due to a different implementation of the wave damping at the outer boundary. 

Using local (shearing sheet) simulations, \citet{Dong2011II} studied vortensity excitation at the planet-driven shock. Their results for $\Delta \zeta (R)$ are compatible with ours regarding the amplitude of the vortensity jump and its radial structure. Their predicted scaling of $\Delta \zeta \propto (M_\mathrm{p}/M_\mathrm{th})^3$ is also consistent with our findings. In our global calculations, we find that the amplitude of $\Delta \zeta$ can sometimes differ by a factor of $\simeq 2-3$ between the inner and outer disc in the most extreme cases, which was not possible for \citet{Dong2011II} to observe due to the symmetry of the shearing sheet setup. 

Both \citet{Dong2011II} and \citet{Duffell2012} found that the azimuthal width of the wake evolves as $\Delta \eta \propto \tau^{1/2}$, translating into a similarly behaving offset of the peak position of the nonlinear wake from the linear prediction (\ref{eq:phi_lin}), in agreement with our findings, see \S\ref{sec:phishock}. Moreover, in their 3D simulations of more massive
planets embedded in discs \citet{Zhu2015} also found good agreement of the nonlinear wake offset with this scaling.


\subsection{Applications of our results}
\label{sect:applications}


Semi-analytical framework for characterizing global planet-driven shocks \citep{Rafikov2002}, further developed and tested in this study, can be applied to improve understanding of various aspects of disc-planet interaction. For example, our results on the shock strength (\S\ref{sec:deltachi}) can be used to compute the contribution of the planet-driven shock heating \citep{Rafikov2016} to thermal balance of the disc; they can also be used for computing the associated mass accretion rate $\dot M(R)$ through the disc. Our semi-analytical fit for the offset between the nonlinear shock location and $\phi_\mathrm{lin}$, see Eq. (\ref{eq:phinonlin}) in \S\ref{sec:phishock}, can be employed for inferring planet masses from the shapes of spirals observed in protoplanetary discs.

Recently \citet{Bollati2021} applied the nonlinear framework of \citetalias{Goodman2001} and \citetalias{Rafikov2002} to model kinematic signatures ("kinks") of planets embedded in the disc, that have been recently observed by {\it ALMA} \citep{Pinte2018,Pinte2019}. They used a procedure identical to the one used in making Fig. \ref{fig:lin-nonlin}c,d --- numerical solution of the Burgers equation (\ref{eq:Burger}) with initial conditions from linear theory --- to compute the velocity perturbation in the disc due to a planet-driven density wave and to obtain the CO emission channel maps. Although we did not compute a kinematic signature in this work, our results (\S\ref{ssec:globwave}) suggest that an approach based on solving Burgers equation may easily overestimate wave decay and underestimate the resultant velocity perturbation and the amplitude of a kink.  

\changed{
While the weakly nonlinear theory of \citetalias{Rafikov2002} strictly applies only to the low planet mass regime, some of its ingredients may be useful also for describing the high-mass ($\Mp\gtrsim \Mth$) regime, relevant for the existing protoplanetary disc observations in CO emission and scattered light. In particular, we expect that, when expressed in terms of the generalized coordinates (\ref{eq:tau_g})-(\ref{eq:chi_g}), the wave properties would still show a universal behavior even for high $\Mp$, which could be calibrated using simulations. 
}

Our results on vortensity generation (\S\ref{sec:resvort}) can be used for understanding the appearance of vortices in planetary vicinity \citep{Koller2003,Li2005,deVal2007}. They can also be employed to (semi-analytically) generate axisymmetric surface density profiles for the planet-induced gaps as a function of time, disc parameters and planet mass, following the recipe of \citet{Lin2010}. These applications will be explored in a forthcoming work
\changed{
(Cimerman \& Rafikov, in prep.).
}
Finally, our results can be useful for benchmarking the codes used for studying disc-planet interaction, especially through diagnostics such as the angular momentum flux evolution and vortensity production at the shock.


\section{Summary}
\label{sec:sum}


We have studied the nonlinear evolution of density waves excited by planets embedded in inviscid, isothermal 2D cylindrical discs with the goal of verifying the weakly nonlinear theory of global density waves developed in  \citet{Rafikov2002}. Using linear calculations, weakly nonlinear theory and full hydrodynamical simulations with Athena++ we explored models for a variety of disc parameters, such as disc scale-heights and surface density slopes, and planet masses spanning two orders of magnitude $M_\mathrm{p}=(0.01-1)M_\mathrm{th}$. Our findings can be briefly summarized as follows.

\begin{itemize}

\item The semi-analytical framework of \citet{Rafikov2002} using rescaled variables $\tau$, $\eta$, $\chi$ (Eqs. \ref{eq:tau_g}-\ref{eq:chi_g}) provides an accurate description of the nonlinear evolution of the global density waves driven by the sub-thermal mass planets, provided that it is properly calibrated using numerical simulations. In particular, many characteristics of the wave exhibit a behaviour close to self-similar when expressed in terms of these variables.

\item We showed that calculation of the density wave properties using Burgers equation (\ref{eq:Burger}), while qualitatively correct, does not provide a quantitatively accurate description of the wave evolution, in particular, overpredicting wave decay, especially for low planet masses. A better approach would be to rely on numerical  calibration of the key inputs of the semi-analytical framework: the global shape of the density wave (Eq. \ref{eq:phinonlin}-\ref{eq:phi-pars}), accounting for the nonlinear effects, and the strength of the shock into which the wave ultimately evolves (Eq. \ref{eq:fit} \& Table \ref{tab:fitpars}).   

\item We derive analytical expressions for the vortensity jump across the global planet-driven shock and verify them using numerical simulations (\S\ref{sec:resvort}). We confirm that vortensity generation at the shock scales as a high power of the planet mass, 
$\Delta \zeta \propto (M_\mathrm{p}/M_\mathrm{th})^3$.

\item Applicability of the weakly nonlinear theory \citep{Rafikov2002} in the disc interior to the planetary orbit is limited by the emergence of secondary spiral arms, which is a linear effect \citep{Bae2018,Miranda2019I}.

\item Our results have implications for understanding the appearance of planet-driven spiral arms in protoplanetary discs, kinematic signatures of embedded planets ("kinks"), and other phenomena.

\end{itemize}

Future work will apply the semi-analytical approach explored in this study to other problems in the area of disc-planet interaction, in particular, the formation of vortices at the edges of planetary gaps driven by the Rossby Wave Instability.

\section*{Acknowledgements}
\textit{Software:} NumPy \citep{2020NumPy-Array}, SciPy \citep{2020SciPy-NMeth}, IPython \citep{IPython}, Matplotlib \citep{Matplotlib}, Athena++ \citep{Athenapp2020}, FARGO3D \citep{FARGO3D} and AstroPy \citep{AstroPyI,AstroPyII}.

We thank the anonymous referee for helpful suggestions, especially in clarifying our explanation of multiple spiral arm formation in the linear case.
We are indebted to Ryan Miranda for many useful exchanges and for sharing his method and code to calculate global linear modes, which greatly enriched this work. We would also like to thank all developers of Athena++ for sharing their code with us and in particular Tomohiro Ono, James Stone and Matt Coleman for helpful discussions. N.P.C. is funded by an Isaac Newton Studentship and a Science and Technology Facilities Council (STFC) studentship. R.R.R. acknowledges financial support through the NASA grant 15-XRP15-2-0139, John N. Bahcall Fellowship, and STFC grant ST/T00049X/1. A large part of the lower resolution simulations were performed on FAWCETT the HPC cluster at DAMTP, University of Cambridge. Special thanks to Kacper Kornet and Deryck Thake for all the support in using local resources. Part of this work was performed using resources provided by the Cambridge Service for Data Driven Discovery (CSD3) operated by the University of Cambridge Research Computing Service (\texttt{www.csd3.cam.ac.uk}), provided by Dell EMC and Intel using Tier-2 funding from the Engineering and Physical Sciences Research Council (capital grant EP/P020259/1), and DiRAC funding from the Science and Technology Facilities Council (\texttt{www.dirac.ac.uk}).

\section*{Data Availability}

The data underlying this article will be shared on reasonable request to the corresponding author.




\bibliographystyle{mnras}
\bibliography{example} 




\appendix


\section{Effect of the form of smoothed potential}
\label{sec:potentialorder}


We tested how our results would change if we replace the $\Phi_\mathrm{p}^{(4)}$ potential (Eq. \ref{eq:phi4}) with a second order smoothed potential of the form
\begin{align}
	\Phi_\mathrm{p}^{(2)} = - \frac{G M_\mathrm{p}}{\left(d^2 + r_\mathrm{s}^2 \right)^{1/2}}.
	\label{eq:plum}
\end{align}
which is commonly employed in modeling planet-disc interaction.

\begin{figure}
\centering
\includegraphics[width=0.49\textwidth]{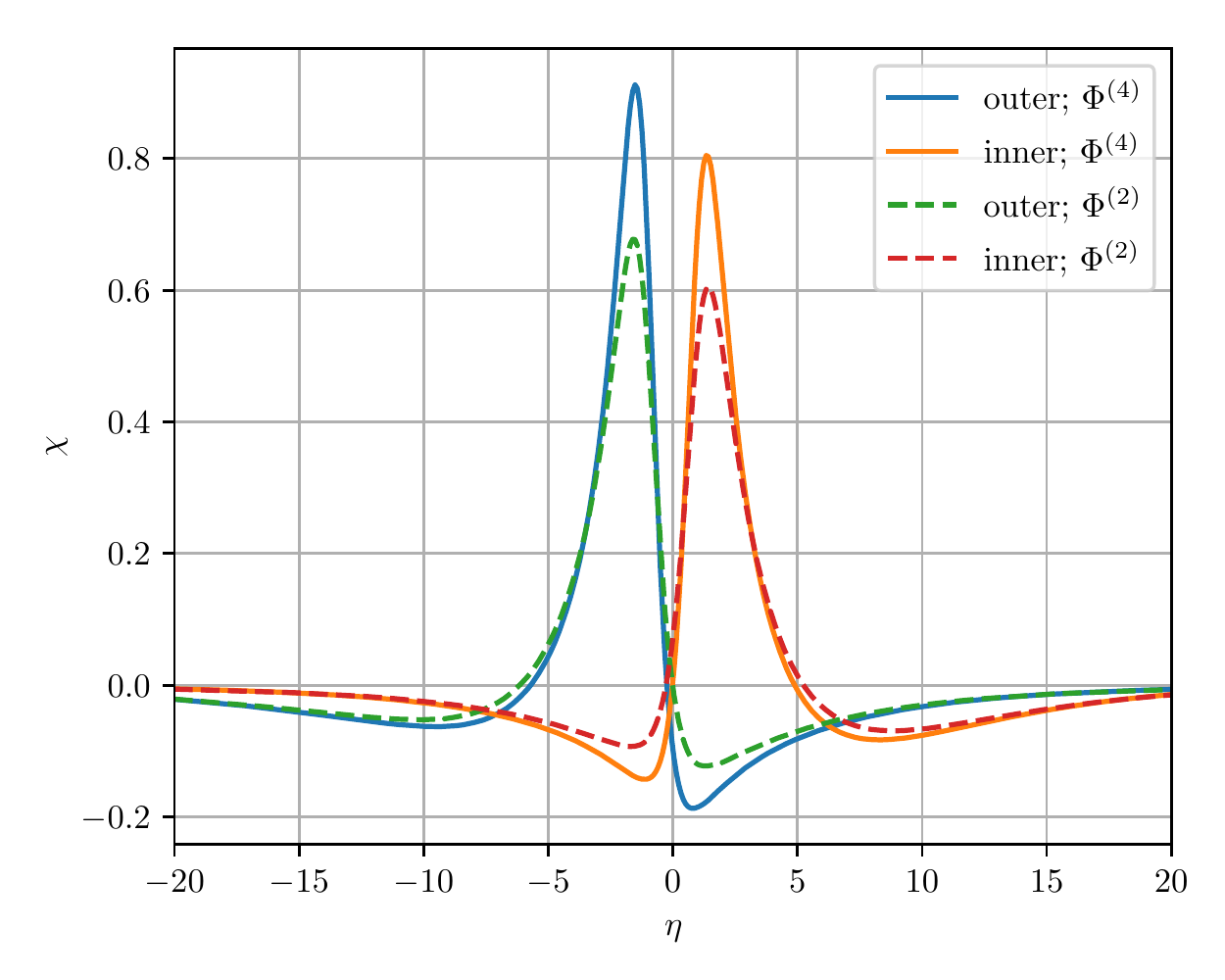}
\vspace*{-0.7cm}
\caption{Initial wake profiles at $|R-R_\mathrm{p}|=1.068R_\mathrm{p}$, $0.935R_\mathrm{p}$ computed using the second and fourth order approximations for the planetary potential, given by the Eqs. (\ref{eq:plum}) and (\ref{eq:phi4}), respectively. Fiducial disc parameters and $M_\mathrm{p} = 0.01 M_\mathrm{th}$ are used.}
\label{fig:2_4_pot}
\end{figure}

The effect on the wave can be seen by examining the wave profile close to the planet. In Fig. \ref{fig:2_4_pot} we show azimuthal wave profiles for the fiducial disc and $M_\mathrm{p} = 0.01 M_\mathrm{th}$ at $R = 1.068 R_\mathrm{p}$ and $R = 0.935 R_\mathrm{p}$. The same smoothing parameter of the gravitational potential, $\epsilon = 0.6$ (typical in the literature, \citealt{Mueller2012}), is used for both potentials.

We find that $\Phi_\mathrm{p}^{(2)}$ results in a wave amplitude that is about 25\% smaller than when using  $\Phi_\mathrm{p}^{(4)}$. This is consistent with the observations of \citet{Dong2011II}, who found that the 2nd order potential creates a shock further from the planet, that is weaker, than $\Phi_\mathrm{p}^{(4)}$. Indeed, as nonlinear effects directly depend on the wave amplitude it is clear that the change of the wave amplitude due to a different potential must affect the time it takes for characteristic to first cross and a shock to be formed. These differences are expected to reduce as the smoothing parameter $\epsilon$ is decreased.


\section{Comparison with another hydro code}
\label{sec:fargo3d}


\begin{figure}
\centering
\includegraphics[width=0.49\textwidth]{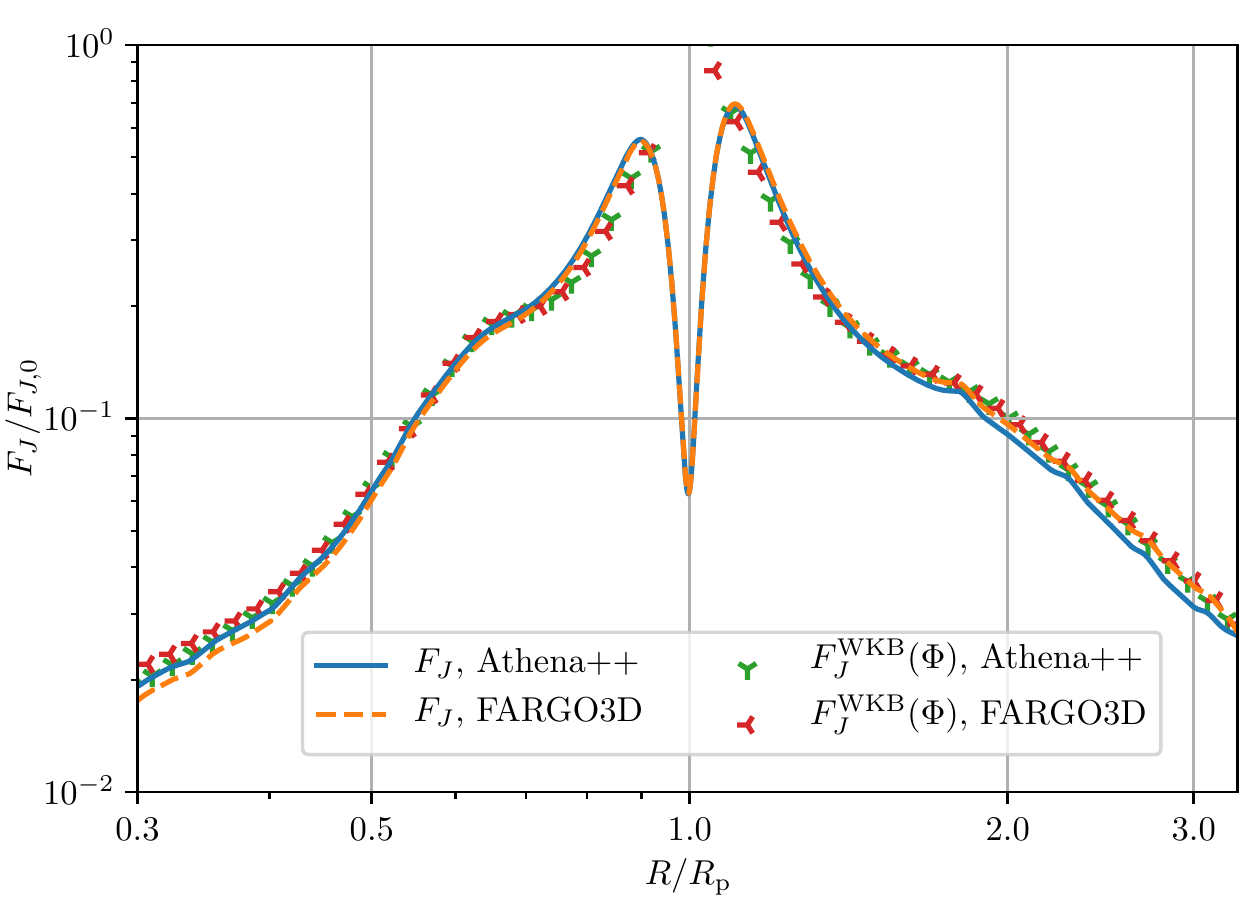}
\vspace*{-0.7cm}
\caption{Comparison of the wave angular momentum fluxes --- curves for $F_J$, dots for $F_J^\mathrm{WKB}$, as shown in the legend --- computed with Athena++ and FARGO3D. Calculation is performed for the same parameters as in Fig. \ref{fig:AMF025}. }
\label{fig:025COMP_FARGO_ATH}
\end{figure}

To test the sensitivity of our results to numerical scheme we also run FARGO3D simulations in order to make a direct comparison with Athena++. We performed simulations at the highest resolution we could achieve on one GPU, $N_R \times N_\phi = 6896 \times 14400$, covering the same domain using the same grid structure, damping zones and boundary conditions as the Athena++ models. For this comparison, we set $\Phi_\mathrm{p}=\Phi_\mathrm{p}^{(4)}$ (Eq. (\ref{eq:phi4})), used the fiducial disc parameters $(q=1.5; h_\mathrm{p}=0.05)$, and show results for $M_\mathrm{p} = 0.25 M_\mathrm{th}$. The behaviour of AMF obtained by both codes is displayed in Fig. \ref{fig:025COMP_FARGO_ATH}. In all regions of the disc we find very good agreement between the two codes. The small disagreement at a few percent level might be due to the lower resolution of FARGO3D runs. 


\section{Derivation of the vortensity jump}
\label{sec:thvort}


We use Eqs. (\ref{eq:g_g}), (\ref{eq:chi_g}) to express the density jump across the shock in a Keplerian disc in terms of the jump of $\chi$ as
\begin{align}
	\frac{\Delta \Sigma}{\Sigma_0} = \frac{M_\mathrm{p}}{M_\mathrm{th}} \left( \sqrt{2} h_\mathrm{p} \frac{\left(R/R_\mathrm{p}\right)^{-p+1}}{\left\vert \left(R/R_\mathrm{p}\right)^{-3/2} - 1 \right\vert} \right)^{1/2} \Delta \chi.
	\label{eq:deltachi}
\end{align}
Given the location of the shock $\phi_\mathrm{sh}(R)$, we can write the derivative along the shock as
\begin{align}
	\td{}{S}	 = \operatorname{sign}(R - R_\mathrm{p}) \times \frac{\de R}{\sqrt{\de R^2 + R^2 \de \phi^2}} \td{}{R} = \frac{\operatorname{sign}(R - R_\mathrm{p})}{\sqrt{1 + R^2 \left(\td{\phi_\mathrm{sh}}{R}\right)^2}} \td{}{R}.
\end{align}
Substituting these results into Eq. (\ref{eq:deltazeta}), we obtain
\begin{align}
	\Delta \zeta = \frac{\cs}{2^{7/4} \Sigma_0 h_\mathrm{p}^{3/2}}  \left( \frac{M_\mathrm{p}}{M_\mathrm{th}} \right)^3  \frac{\left\vert \left(R/R_\mathrm{p}\right)^{-3/2} - 1 \right\vert}{\left(R/R_\mathrm{p}\right)^{-p+1}} \left(\Delta \chi\right)^2 \label{eq:thvort} \\ \nonumber
	\times \left[1 + \frac{M_\mathrm{p}}{M_\mathrm{th}} \left( \frac{\left\vert \left(R/R_\mathrm{p}\right)^{-3/2} - 1 \right\vert}{\sqrt{2} h_\mathrm{p} \left(R/R_\mathrm{p}\right)^{-p+1}} \right)^{1/2} \Delta \chi \right]^{-5/2} \\ \nonumber
	\times \frac{\operatorname{sign}(R - R_\mathrm{p})}{\sqrt{1 + R^2 \left(\td{\phi_\mathrm{sh}}{R}\right)^2}} \td{}{R} \left[ \left( \frac{\left\vert \left(R/R_\mathrm{p}\right)^{-3/2} - 1 \right\vert}{\left(R/R_\mathrm{p}\right)^{-p+1}} \right)^{1/2} \Delta \chi \right]
\end{align}
The factor in the middle line of the above expression is $\M^{-5}$, where $\M$ is the normal Mach number of the shock.
For $M_\mathrm{p}\lesssim M_\mathrm{th}$ it is typically of order unity. Defining
\begin{align}
	B(R) &\equiv \left[\left(R/R_\mathrm{p}\right)^{p-1} \left\vert \left(R/R_\mathrm{p}\right)^{-3/2} - 1 \right\vert\right]^{1/2},
	\label{eq:BR}\\
	C(R) &\equiv \operatorname{sign}(R - R_\mathrm{p})\left[1 + R^2 \left(\td{\phi_\mathrm{sh}}{R}\right)^2\right]^{-1/2},
	\label{eq:CR}
\end{align}
Eq. (\ref{eq:thvort}) can be written more compactly as Eq. (\ref{eq:deltazeta2}).

Information about the shape of the shock enters Eq. (\ref{eq:deltazeta2}) through the factor $C(R)$.  When using the linear prediction for the location of the primary arm, Eq. (\ref{eq:philin}), one finds
\begin{align}
	C(R)	 = \frac{\operatorname{sign}(R - R_\mathrm{p})}{\sqrt{1 + h_\mathrm{p}^{-2} \left(\frac{R}{R_\mathrm{p}}\right)^2\left[ \left(\frac{R}{R_\mathrm{p}}\right)^{-3/2} - 1 \right]^2}}.
	\label{eq:C-lin}
\end{align}
Alternatively, when accounting for the nonlinear correction to the shock position, Eq. (\ref{eq:phinonlin}), one gets
\begin{align}
	C(R)	 = \frac{\operatorname{sign}(R - R_\mathrm{p})}{\sqrt{1 +h_\mathrm{p}^{-2} \left(\frac{R}{R_\mathrm{p}}\right)^2\left[ \left(\frac{R}{R_\mathrm{p}}\right)^{-3/2} - 1 + \frac{\Delta\phi_0 h_\mathrm{p}^2 R_\mathrm{p}}{2(\tau-\tau_0)^{1/2}} \td{\tau}{R}\right]^2}}.
	\label{eq:C-nonlin}
\end{align}
The nonlinear correction acts to decrease the curvature of the shock and reduces the denominator in Eq. (\ref{eq:C-nonlin}), effectively increasing the derivative along the shock $\mathrm{d}/\mathrm{d}S$.


\bsp	
\label{lastpage}
\end{document}